\documentclass[reprint,twocolumn]{revtex4-2}
\usepackage{amsfonts}
\usepackage{amsmath}
\usepackage{amssymb}
\usepackage{charter}
\usepackage{graphicx}
\usepackage{float}
\usepackage{amsmath}
\usepackage{amssymb}
\usepackage{charter}
\usepackage{graphicx}
\usepackage{subfigure}
\usepackage{graphicx}
\usepackage{epstopdf}
\usepackage{color}
\usepackage{tocvsec2}
\usepackage{enumerate}
\usepackage{graphicx}
\usepackage{dcolumn}
\usepackage{bm}

\begin{document}

\title{ Phase estimation via number-conserving operation inside the SU(1,1)
interferometer }
\author{Qingqian Kang$^{1,2}$}
\author{Zekun Zhao$^{1}$}
\author{Teng Zhao$^{1}$}
\author{Cunjin Liu$^{1}$}
\author{Liyun Hu$^{1,3}$}
\thanks{hlyun@jxnu.edu.cn}
\affiliation{$^{{\small 1}}$\textit{Center for Quantum Science and Technology, Jiangxi
Normal University, Nanchang 330022, China}\\
$^{{\small 2}}$\textit{Department of Physics, Jiangxi Normal University Science and
Technology College, Nanchang 330022, China}\\
$^{{\small 3}}$\textit{Institute for Military-Civilian Integration of Jiangxi Province,
 Nanchang 330200, China}}

\begin{abstract}
Utilizing nonlinear elements, SU(1,1) interferometers demonstrate superior
phase sensitivity compared to passive interferometers. However, the
precision is significantly impacted by photon losses, particularly internal
losses. We propose a theoretical scheme to improve the precision of phase
measurement using homodyne detection by implementing number-conserving
operation (NCO), i.e., $aa^{\dagger }$ and $a^{\dagger }a$, inside the
SU(1,1) interferometer, with the coherent state and the vacuum state as the
input states. We analyze the effects of NCO on the phase sensitivity, the
quantum Fisher information (QFI), and the quantum Cram\'{e}r-Rao bound
(QCRB) under both ideal and photon losses scenarios. Our findings reveal
that the internal non-Gaussian operations can enhance the phase sensitivity
and the QFI, and effectively improve the robustness of the SU(1,1)
interferometer against internal photon losses. Notably, the $a^{\dagger }a$
scheme exhibits superior improvement in both ideal and photon losses cases
in terms of phase sensitivity. Moreover, in the ideal case, $aa^{\dagger }$
scheme slightly outperforms $a^{\dagger }a$ scheme in terms of the QFI.
However, in the presence of high photon losses, $a^{\dagger }a$ scheme
demonstrates a greater advantage.

\textbf{PACS: }03.67.-a, 05.30.-d, 42.50,Dv, 03.65.Wj
\end{abstract}

\maketitle

\section{Introduction}

Optical interference measurement plays a crucial role in many scientific and
technological applications such as quantum metrology for precise
measurements, imaging for capturing detailed visual information, sensing for
detecting and measuring physical quantities, and information processing for
manipulating and transmitting data \cite{a1,a2,a3,a4,a5,a6,a7,a8,a9}.
Consequently, there has been extensive research and significant advancements
in the field of optical interference measurement. To satisfy the need for
high precision, a variety of optical interferometers have been proposed and
developed. One of the most practical interferometers is the Mach-Zehnder
interferometer (MZI), whose phase sensitivity is limited by the standard
quantum-noise limit (SQL) $\Delta \phi =1/\sqrt{N}$ ($N$ is the average
number of photons within the interferometer), together with solely classical
resources as the input of the MZI \cite{a10}. Over recent decades, various
schemes have been proposed to improve the phase sensitivity of the
traditional MZI \cite{b1,b2}. It has been demonstrated that the quantum
states as the input states to make the traditional MZI beat the SQL. For
example, NOON state \cite{b3,b4}, twin Fock state \cite{b5}, and the
squeezed state \cite{b7,b71} \textit{et al.} can achieve or even exceed the
Heisenberg limit (HL) $\Delta \phi =1/N$ \cite{b9,b91}.

Another possibility to realize quantum-enhanced phase sensitivity is the
SU(1,1) interferometer \cite{b10,b11}, which replaced traditional linear
beam splitters (BSs) with optical parametric amplifiers (OPAs). It splits
and mixes beams using nonlinear transformations, which is first proposed by
Yurke \textit{et al.} \cite{b12}. In the SU(1,1) interferometer comprising
two OPAs, the first OPA serves the dual purpose of acquiring entangled
resources and suppressing amplified noise. Meanwhile, the subsequent use of
the second OPA can lead to signal enhancement, offering a viable pathway for
achieving higher precision in phase estimation. By utilizing entangled
photon states, the SU(1,1) interferometer can surpass the SQL, enabling
higher precision. This technique revolutionized phase estimation, becoming a
vital tool in quantum precision measurements. Then, there has been
significant interest in studying the SU(1,1) interferometer \cite%
{b13,b14,b15}. For instance, Hudelist \textit{et al.} demonstrated that the
gain effect of OPA results in the SU(1,1) interferometer exhibiting higher
sensitivity compared to traditional linear interferometers \cite{b16}. In
2011, Jing \textit{et al.} \cite{b17} successfully implemented this
interferometer experimentally. In this nonlinear interferometer, the maximum
output intensity can be much higher than that of linear interferometer due
to the OPA. Apart from the standard form, various configurations of SU(1,1)
interferometer have also been proposed \cite{b14,m1,m2,m3,m4,m5,m6,m7,m8,m9}.

As previously mentioned, although SU(1,1) interferometer is highly valuable
for precision measurement \cite{c1,c1a}, the precision is still affected by
dissipation, particularly photon losses inside the interferometer \cite%
{c2,c3}. Consequently, to further enhance precision, non-Gaussian operations
should serve as an effective approach to mitigate internal dissipation. Most
theoretical \cite{c4,c5,c6,c7} and experimental \cite{c8,c9,c10} studies
have fortunately indicated that non-Gaussian operations, such as photon
subtraction (PS), photon addition (PA), photon catalysis (PC), quantum
scissor and their coherent superposition, are effectively enhancing the
nonclassicality and entanglement degrees of quantum states, thereby
enhancing their potential in quantum information processing \cite{c11,c111}.
Experimental studies have illustrated the conditional generation of
superpositions of distinct quantum operations through single-photon
interference, providing a practical approach for preparing non-Gaussian
operations \cite{c12}.\textbf{\ }This advancement has unveiled new
possibilities in quantum state manipulation and implications for various
quantum technologies. In Ref. \cite{c121}, Zhang et al. proposed a
number-conserving operation (NCO) on the inputs of MZI for improving the
resolution and precision of phase measurement with parity detection. It is
shown that the phase sensitivity can be better than that of both the photon
subtraction operation and photon addition operation, in the presence of
photon losses. Different from Ref. \cite{c121}, Xu et al. examined the phase
sensitivity with internal photon losses in SU(1,1) interferometers, rather
than in the MZI, an SU(2) interferometer. It is found that performing photon
addition operations internally provides superior results compared to those
at the input \cite{c13}. Thus, a question arises naturally: can the NCO be
operated inside the SU(1,1) interferometer (i.e., the non-Gaussian operation
on the output states after first OPA) to mitigate the impact of internal
photon losses?

Therefore, in this paper, we concentrate on employing the NCO scheme inside
the SU(1,1) interferometer to enhance the measurement accuracy , and then
analyze the improvement effect of internal non-Gaussian operation on the
phase sensitivity and the QFI in the presence of photon losses. The
remainder of this paper is arranged as follows. Sec. II outlines the
theoretical model of the NCO. Sec. III delves into phase sensitivity,
encompassing both ideal and internal photon losses cases. Sec. IV centers on
the QFI and QCRB \cite{c14,c15}. Finally, Sec. V provides a comprehensive
summary.

\section{Model}

This section begins with an introduction to the SU(1,1) interferometer, as
illustrated in Fig. 1(a). The SU(1,1) interferometer typically consists of
two OPAs and a linear phase shifter, making it one of the most commonly used
interferometers in quantum metrology research. The first OPA is
characterized by a two-mode squeezing operator $U_{S_{1}}(\xi _{1})=\exp
(\xi _{1}^{\ast }ab-\xi _{1}a^{\dagger }b^{\dagger })$, where $a$ and $b$, $%
a^{\dagger }$ and $b^{\dagger }$ represent the photon annihilation and
creation operators, respectively. The squeezing parameter $\xi _{1}$ can be
expressed as $\xi _{1}=g_{1}e^{i\theta _{1}}$, where $g_{1}$ represents the
gain factor and $\theta _{1}$ represents the phase shift. This parameter
plays a critical role in shaping the interference pattern and determining
the system's phase sensitivity. Following the first OPA, mode $a$ undergoes
a phase shift process $U_{\phi }=\exp [i\phi (a^{\dagger }a)]$, while mode $%
b $ remains unchanged. Subsequently, the two beams are coupled in the second
OPA with the operator $U_{S_{2}}(\xi _{2})=\exp (\xi _{2}^{\ast }ab-\xi
_{2}a^{\dagger }b^{\dagger })$, where $\xi _{2}=g_{2}e^{i\theta _{2}}$ and $%
\theta _{2}-\theta _{1}=\pi $. In this paper, we set the parameters $%
g_{1}=g_{2}=g$, $\theta _{1}=0$, $\theta _{2}=\pi $. We utilize the coherent
state $\left \vert \alpha \right \rangle _{a}$ and the vacuum state $\left
\vert 0\right \rangle _{b}$ as input states and homodyne detection is
employed on the mode $a$ of the output.

The SU(1,1) interferometer is generally susceptible to photon losses,
particularly in the case of internal losses. To simulate photon losses, the
use of fictitious BSs is proposed, as depicted in Fig. 1(a). The operators
of these fictitious BSs can be represented as $U_{B}=U_{B_{a}}\otimes
U_{B_{b}}$, with $U_{B_{a}}=\exp \left[ \theta _{a}\left( a^{\dagger
}a_{v}-aa_{v}^{\dagger }\right) \right] $ and $U_{B_{b}}=\exp \left[ \theta
_{b}\left( b^{\dagger }b_{v}-bb_{v}^{\dagger }\right) \right] $, where $%
a_{v} $ and $b_{v}$ represent vacuum modes. Here, $T_{k}$ ($k=a,b$) denotes
the transmissivity of the fictitious BSs, associated with $\theta _{k}$
through $T_{k}=\cos ^{2}\theta _{k}\in \left[ 0,1\right] $. The value of
transmittance equal to $1$ ($T_{k}=1$) corresponds to the ideal case without
photon losses \cite{c13}. In an expanded space, the expression for the
output state of the standard SU(1,1) interferometer can be represented as
the following pure state, i.e.,%
\begin{equation}
\left \vert \Psi _{out}^{0}\right \rangle =U_{S_{2}}U_{\phi
}U_{B}U_{S_{1}}|\psi _{in}\rangle ,  \label{eq1}
\end{equation}%
where $|\psi _{in}\rangle =$ $\left \vert \alpha \right \rangle
_{a}\left
\vert 0\right \rangle _{b}\left \vert 0\right \rangle
_{a_{v}}\left \vert 0\right \rangle _{b_{v}}$.

\begin{figure}[tph]
\label{Figure1} \centering \includegraphics[width=0.9\columnwidth]{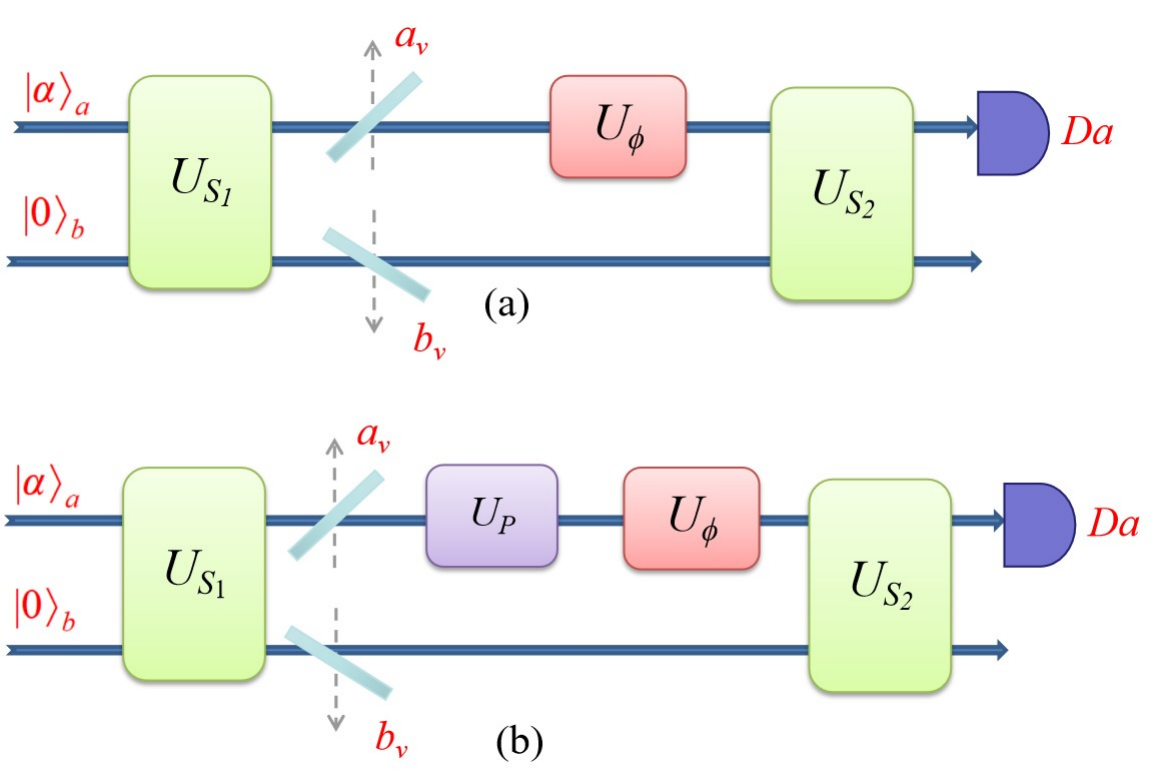}
\
\caption{Schematic diagram of the SU(1,1) interferometer. (a) the standard
SU(1,1) interferometer, (b) the SU(1,1) interferometer with NCO. The two
input ports are a coherent state $\left \vert \protect \alpha \right \rangle
_{a}$ and a vacuum state $\left \vert 0\right \rangle _{b}$. $a_{v}$ and $%
b_{v} $ are vacuum modes. $U_{S_{1}}$ and $U_{S_{2}}$\ are the optical
parametric amplifiers, $U_{\protect \phi }$ is the phase shifter. $U_{P}$ is
the NCO operator and $D_{a}$ is the homodyne detector.}
\end{figure}

To mitigate the impact of photon losses, we introduce a distinct
non-Gaussian operation inside the SU(1,1) interferometer, called the NCO
scheme, as illustrated in Fig. 1(b). We utilize simple and easy-to-prepare
input states ( $\left \vert \alpha \right \rangle _{a}\otimes \left \vert
0\right \rangle _{b}$ ), and an experimentally feasible homodyne detection.
As referred to Ref. \cite{c7}, the NCO can be seen as an equivalent
operator, i.e.,
\begin{equation}
U_{P}=saa^{\dagger }+ta^{\dagger }a,  \label{eq2}
\end{equation}%
where $s^{2}+t^{2}=1$ with $s$\ and $t$\ being real numbers, $a$ and $%
a^{\dagger }$ are annihilation operator and creation operator, respectively.
From Eq. (\ref{eq2}), one can obtain the photon addition then\ photon
subtraction\ (PA-then-PS) $aa^{\dagger }$, and photon subtraction then
photon addition (PS-then-PA) $a^{\dagger }a$, respectively. The process can
be described by operator $U_{P_{j}}$, where $j=1$ and $2$, $%
U_{P_{1}}=aa^{\dagger }$ and $U_{P_{2}}=a^{\dagger }a$, respectively.
Actually, the NCOs $aa^{\dagger }$\ and $a^{\dagger }a$\ are non-Gaussian
operations, which can be experimentally realized via conditional
measurement, like $a$\ and $a^{\dagger }$. For instance, on the basis of BS
with high transmissivity and a photon detection, one can arrive at the
experimental implementation of a single PS \cite{c10}. In addition, PA
operation can be implemented by four-wave mixing technique, and it is also
implemented by BS with zero photon detection and single photon input \cite%
{c15a,c15b}. When the two consecutive conditional measurements are achieved,
the quantum state corresponding to the detected results is selected to be
our study object. In the ideal case, the obtained state is not a mixed state
but a pure one for a pure input.

In this case of NCO applying inside the SU(1,1) interferometer, the output
state can be written as the following pure states
\begin{equation}
\left \vert \Psi _{out}^{1}\right \rangle =A_{1}U_{S_{2}}U_{\phi
}U_{p_{_{1}}}U_{B}U_{S_{1}}|\psi _{in}\rangle ,  \label{eq3}
\end{equation}%
and
\begin{equation}
\left \vert \Psi _{out}^{2}\right \rangle =A_{2}U_{S_{2}}U_{\phi
}U_{p_{_{2}}}U_{B}U_{S_{1}}|\psi _{in}\rangle .  \label{eq4}
\end{equation}%
$A_{1}$ and $A_{2}$\ are the normalization constants for the PA-then-PS and
PS-then-PA, respectively, given by \cite{c13}
\begin{eqnarray}
A_{1} &=&\left( P_{2,2,0,0}+3P_{1,1,0,0}+1\right) ^{-\frac{1}{2}},
\label{eq6} \\
A_{2} &=&\left( P_{2,2,0,0}+P_{1,1,0,0}\right) ^{-\frac{1}{2}},  \label{eq7}
\end{eqnarray}%
where $P_{x_{1},y_{1},x_{2},y_{2}}=\partial
^{x_{1}+y_{1}+x_{2}+y_{2}}/\partial \lambda _{1}^{x_{1}}\partial \lambda
_{2}^{y_{1}}\partial \lambda _{3}^{x_{2}}\partial \lambda _{4}^{y_{2}}$ $%
\left. \left \{ e^{w_{4}}\right \} \right \vert _{\lambda _{1}=\lambda
_{2}=\lambda _{3}=\lambda _{4}=0},$ as well as%
\begin{eqnarray}
w_{1} &=&\lambda _{1}T\left( \lambda _{2}\sinh g-\lambda _{3}\cosh g\right)
\sinh g  \notag \\
&&+\lambda _{4}T\left( \lambda _{3}\sinh g-\lambda _{2}\cosh g\right) \sinh
g,  \label{eq8} \\
w_{2} &=&\lambda _{1}\sqrt{T}\cosh g-\lambda _{4}\sqrt{T}\sinh g,\text{ }
\label{eq9} \\
\text{\  \  \  \ }w_{3} &=&\lambda _{2}\sqrt{T}\cosh g-\lambda _{3}\sqrt{T}%
\sinh g,  \label{eq10} \\
w_{4} &=&w_{1}+w_{2}\alpha ^{\ast }+w_{3}\alpha .  \label{eq11}
\end{eqnarray}

\section{Phase sensitivity}

Quantum metrology is an effective approach utilizing quantum resources for
precise phase measurements \cite{d1,d2}. The objective is to achieve highly
sensitive measurements of unknown phases. Within this section, we delve
further into investigating the phase sensitivity for the NCO inside the
SU(1,1) interferometer \cite{d3}. Various detection methods are available
for this purpose, such as homodyne detection \cite{d4,d5}, parity detection
\cite{b7,d7}, and intensity detection \cite{d8}. Each of these methods
offers different trade-offs between sensitivity, complexity, and practical
implementation. It is important to note that the phase sensitivities of
different detection schemes may vary for different input states and
interferometers \cite{d9}. Each measurement method has its own advantages.
In many schemes, parity detection has been proven to be the optimal
detection method for linear phase estimation \cite{b7,b91}, but it is
relatively complex and is harder to implement experimentally. In Ref. \cite%
{d4}, it is noted that the phase sensitivity of an SU(1,1) interferometer
with homodyne detection surpasses that with intensity detection. Homodyne
detection is not only easy to implement with current experimental technology
\cite{c15a}, but also simple from a theoretical calculation perspective,
thereby playing a key role in the field of continuous-variable quantum key
distribution \cite{c16,c17}. For this reason, we use the homodyne detection
on mode $a$\ at one output port to estimate the phase sensitivity.

In homodyne detection, the measured variable is one of the two orthogonal
components of the mode $a$, given by $X=(a+a^{\dagger })/\sqrt{2}$. Based on
the error propagation equation \cite{b12}, the phase sensitivity can be
expressed as
\begin{equation}
\Delta \phi =\frac{\sqrt{\left \langle \Delta ^{2}X\right \rangle }}{%
|\partial \left \langle X\right \rangle /\partial \phi |}=\frac{\sqrt{\left
\langle X^{2}\right \rangle -\left \langle X\right \rangle ^{2}}}{|\partial
\left \langle X\right \rangle /\partial \phi |}.  \label{eq12}
\end{equation}

Based on Eqs. (\ref{eq3}), (\ref{eq4}) and (\ref{eq12}), the phase
sensitivity for the NCO can be theoretically determined. The detail
calculation steps for the phase sensitivity $\Delta \phi $ of the PA-then-PS
and PS-then-PA are provided in Appendix A.

\subsection{Ideal case}

Initially, we consider the ideal case, $T_{k}=1$ (where $k=a,b$),
representing the scenario without photon losses. The phase sensitivity $%
\Delta \phi $ is plotted as a function of $\phi $ in Fig. 2. Fig. 2(a) is
for different superposition operations, from which it is observed that when $%
0<t<1$ (dashed line), the phase sensitivity consistently falls between the
extremes of $t=0$ (red solid line) and $t=1$ (blue solid line), which
correspond to PA-then-PS and PS-then-PA, respectively. This indicates that
the effects of superposition operations are between PS-then-PA and
PA-then-PS on the improvement of phase sensitivity. Thus, for the sake of
simplicity, our subsequent investigation concentrates only on these two
boundary cases. Fig. 2(b) is for the standard and these two boundary cases.
It is shown that (i) The phase sensitivity improves initially and then
decreases as the phase increases, with the optimal sensitivity deviating
from $\phi =0$. (ii) Both PA-then-PS and PS-then-PA effectively enhance the
phase sensitivity $\Delta \phi $. (iii) The phase sensitivity of PS-then-PA
is better than that of PA-then-PS, and the difference increases with
increasing phase.

\begin{figure}[tph]
\label{Figure2} \centering%
\subfigure{
\begin{minipage}[b]{0.5\textwidth}
\includegraphics[width=0.83\textwidth]{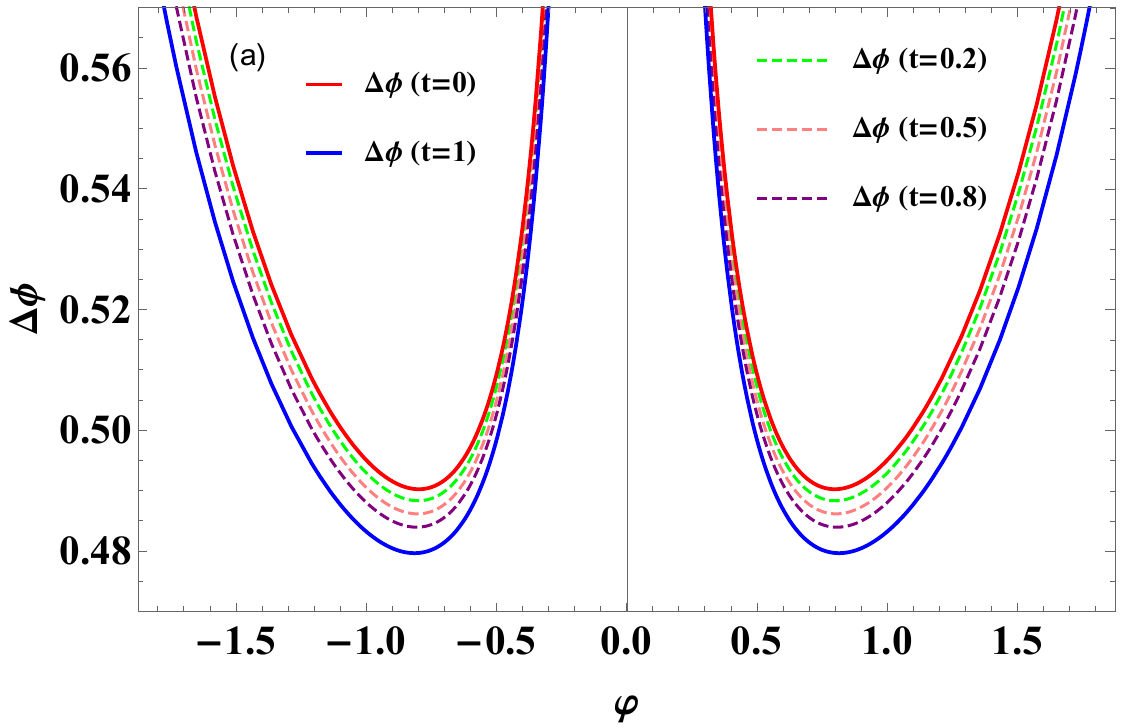}\\
\includegraphics[width=0.83\textwidth]{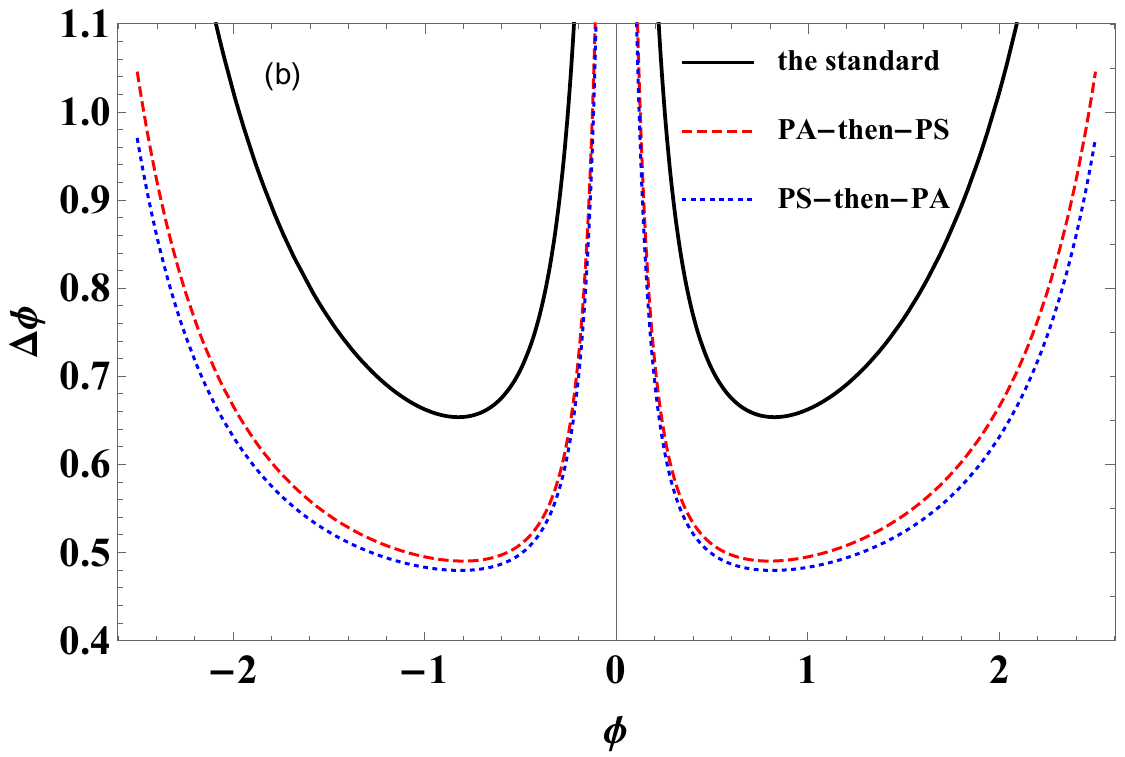}
\end{minipage}}
\caption{The phase sensitivity of NCO based on the homodyne detection as a
function of $\protect \phi $ with $\protect \alpha =1$ and $g=1$. (a) The
phase sensitivity for different values of the parameter\ $t$. (b) The black
solid line corresponds to the standard SU(1,1) interferometer; the red
dashed line and the blue dotted line correspond to the PA-then-PS and
PS-then-PA, respectively.}
\end{figure}

Fig. 3 illustrates that the phase sensitivity $\Delta \phi $ plotted against
the gain factor $g$ for different schemes. The plot confirms that an
increase in the gain factor $g$ enhances the phase sensitivity. It is
interesting to notice that, when the $g$ value is small, the PA-then-PS
demonstrates a better improvement. Conversely, when the $g$ value is large,
the opposite is true. Although the improvement of both is related to
parameter $g$, the PS-then-PA is better in terms of accuracy, i.e., the
PS-then-PA achieves the optimal phase sensitivity. Thus, the following
studies mainly focus on the large parameter $g$.
\begin{figure}[tph]
\label{Figure3} \centering \includegraphics[width=0.9\columnwidth]{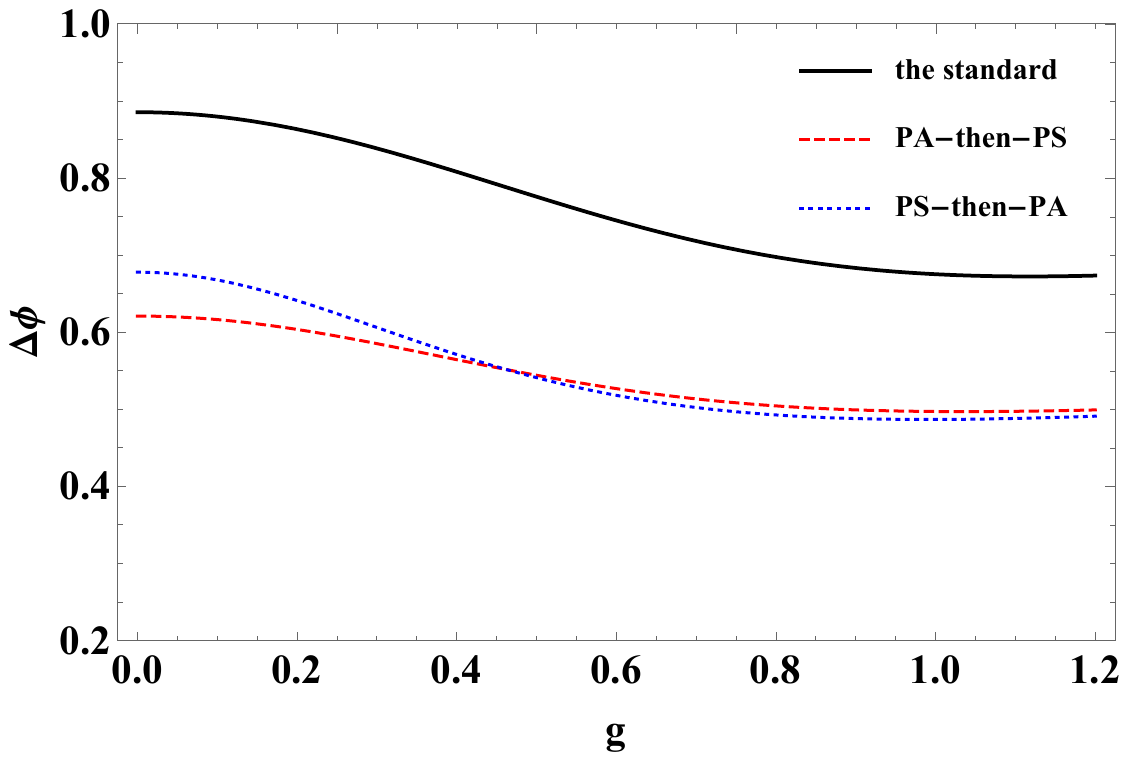}
\
\caption{The phase sensitivity as a function of $g$, with $\protect \alpha %
=1\ $and $\protect \phi =0.6$. }
\end{figure}

Similarly, we analyze the phase sensitivity $\Delta \phi $ as a function of
the coherent amplitude $\alpha $, as depicted in Fig. 4. The phase
sensitivity improves with the coherent amplitude $\alpha $, attributed to
the increase in the mean photon number with $\alpha $, then enhancing
intramode correlations and quantum entanglement between the two modes.
Furthermore, the enhancement effect diminishes as the coherent amplitude $%
\alpha $ increases. It is noteworthy that the PS-then-PA demonstrates better
improvement than the PA-then-PS at small values of $\alpha $, while the
improvement effects of both schemes are consistent at larger values of $%
\alpha $.

\begin{figure}[tph]
\label{Figure4} \centering \includegraphics[width=0.9\columnwidth]{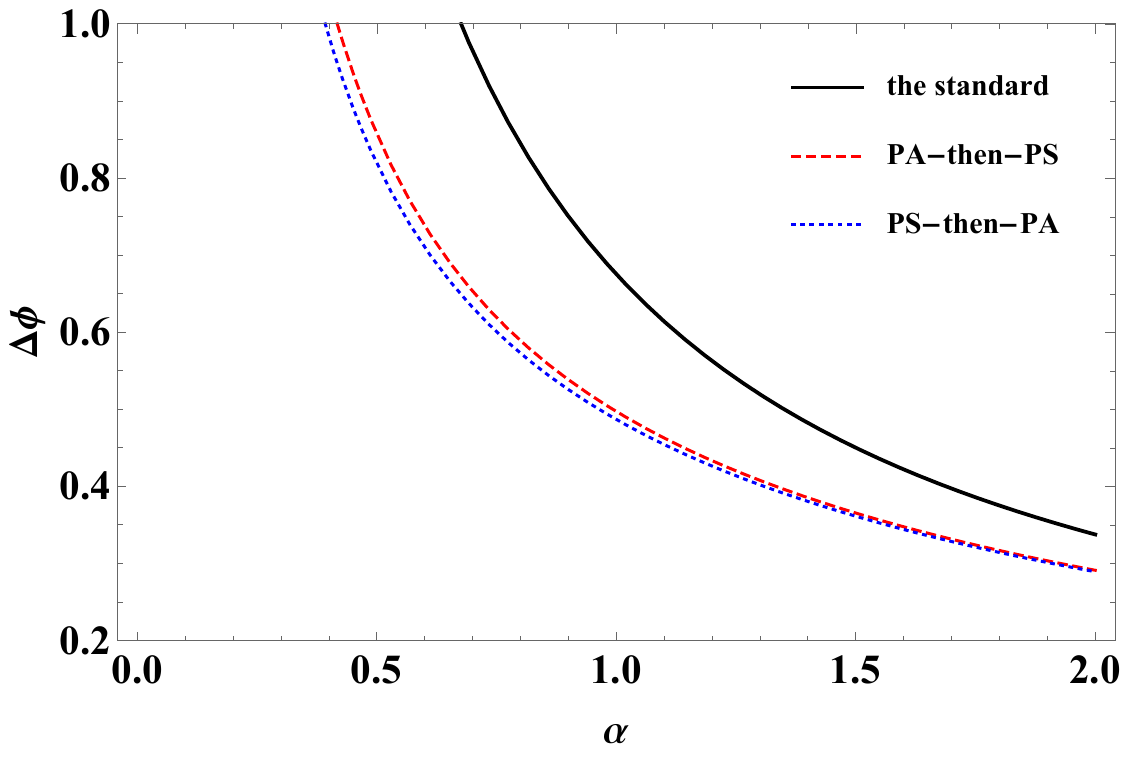}
\
\caption{The phase sensitivity as a function of $\protect \alpha $, with $%
g=1\ $and $\protect \phi =0.6$. }
\end{figure}

\subsection{Photon losses case}

The SU(1,1) interferometer plays a critical role in achieving high-precision
measurements. However, precision is significantly affected by photon losses,
particularly internal losses. Here, we focus on internal photon losses,
corresponding to $T_{k}\in (0,1)$. The phase sensitivity, depicted as a
function of transmittance $T_{k}$ in Fig. 5 for fixed $g$, $\alpha $, and $%
\phi $, improves as anticipated with higher transmittance $T_{k}$. Lower
transmittance corresponds to increased internal losses, weakening the
performance of phase estimation. Both PA-then-PS and PS-then-PA schemes
within the SU(1,1) interferometer effectively enhance the phase sensitivity $%
\Delta \phi $. Moreover, it is notable that as transmittance $T_{k}$
increases, the improvement in phase sensitivity first increases and then
decreases for both schemes. Notably, the PS-then-PA scheme consistently
demonstrates higher phase sensitivity than the PA-then-PA scheme across the
entire range.

\begin{figure}[tph]
\label{Figure5} \centering \includegraphics[width=0.9\columnwidth]{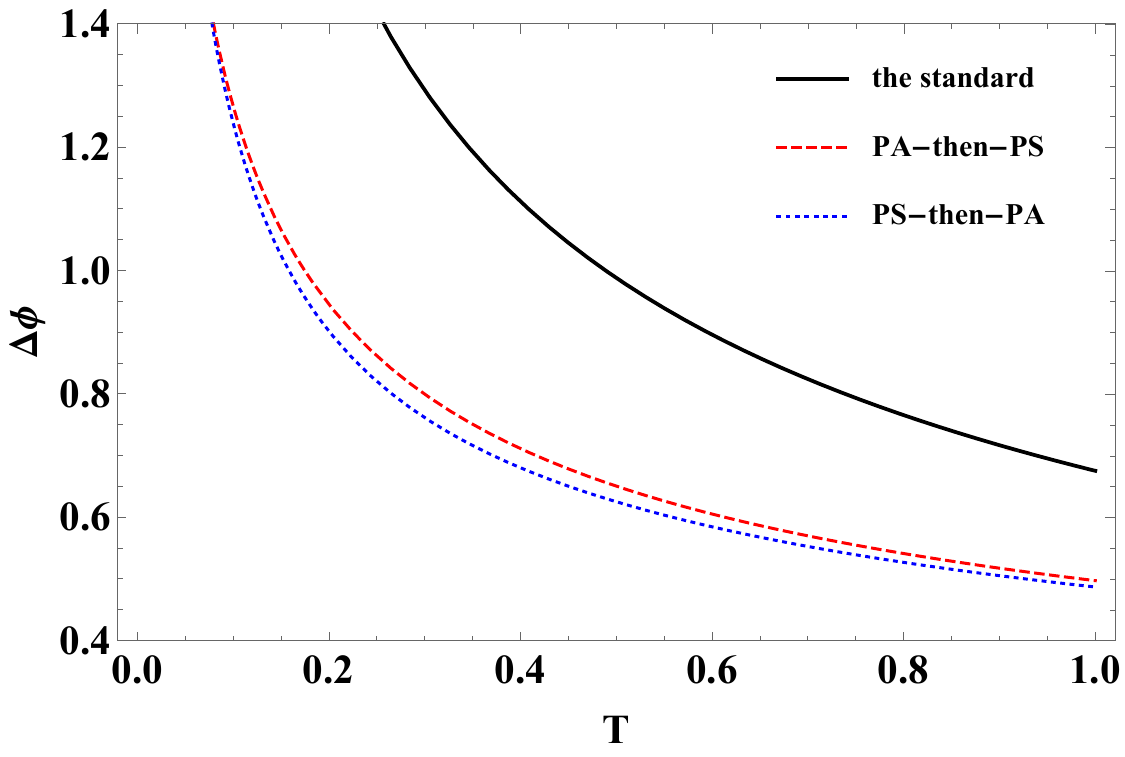}%
\
\caption{The phase sensitivity as a function of transmittance $T_{k}$, with $%
g=1$, $\protect \phi =0.6$ and $\protect \alpha =1.$ }
\end{figure}

The robustness to photon losses denotes the measurement process's
insensitivity to photon losses. A quantum precision measurement system with
strong robustness can maintain high accuracy and stability even in the
presence of photon losses, thereby reducing measurement errors and
uncertainties. By designing and optimizing interferometer measurement
processes, the system's robustness to photon losses can be improved.

To better study the enhancing effect of the NCO on robustness against photon
losses, we further compare the changes of the phase sensitivity in ideal and
photon losses cases for different schemes, as shown in Fig. 6. The
comparison reveals that the phase sensitivity of the standard SU(1,1)
interferometer is more significantly affected by photon losses. In contrast,
the phase sensitivity of the NCO is less affected, indicating that the
non-Gaussian operations can mitigate the impact of internal photon losses
and enhance the interferometer's robustness against losses.

\begin{figure}[tph]
\label{Figure6} \centering%
\subfigure{
\begin{minipage}[b]{0.5\textwidth}
\includegraphics[width=0.83\textwidth]{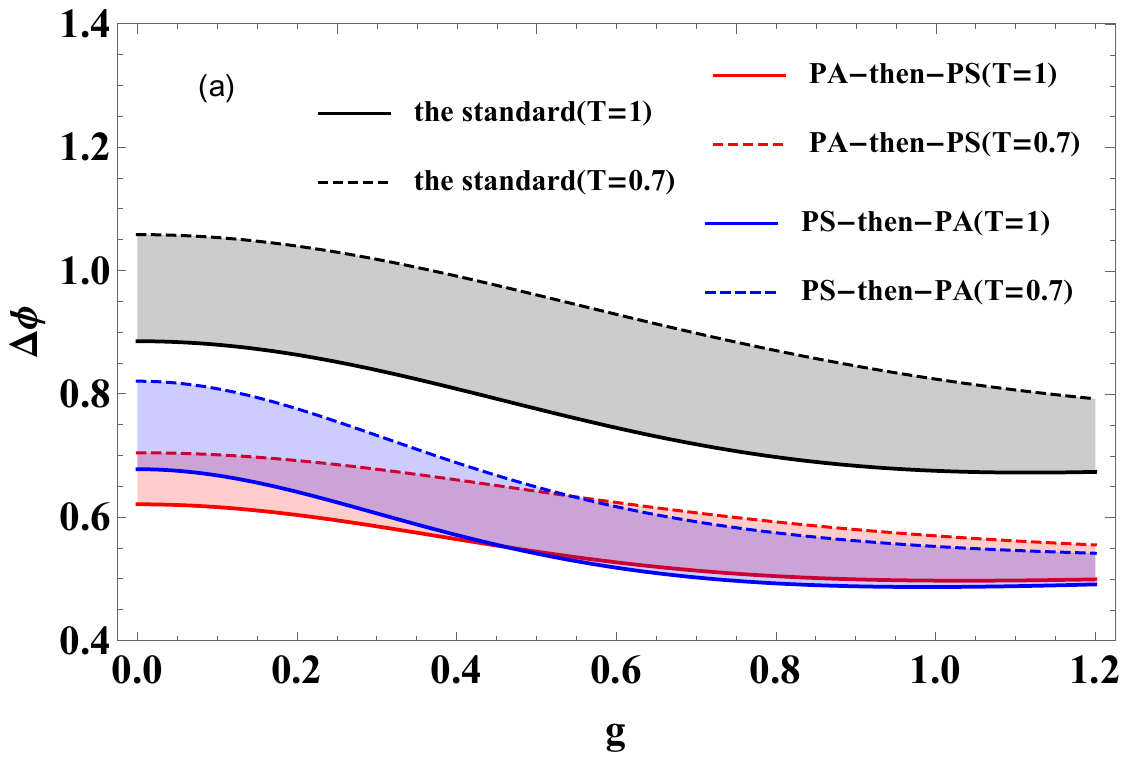}\\
\includegraphics[width=0.83\textwidth]{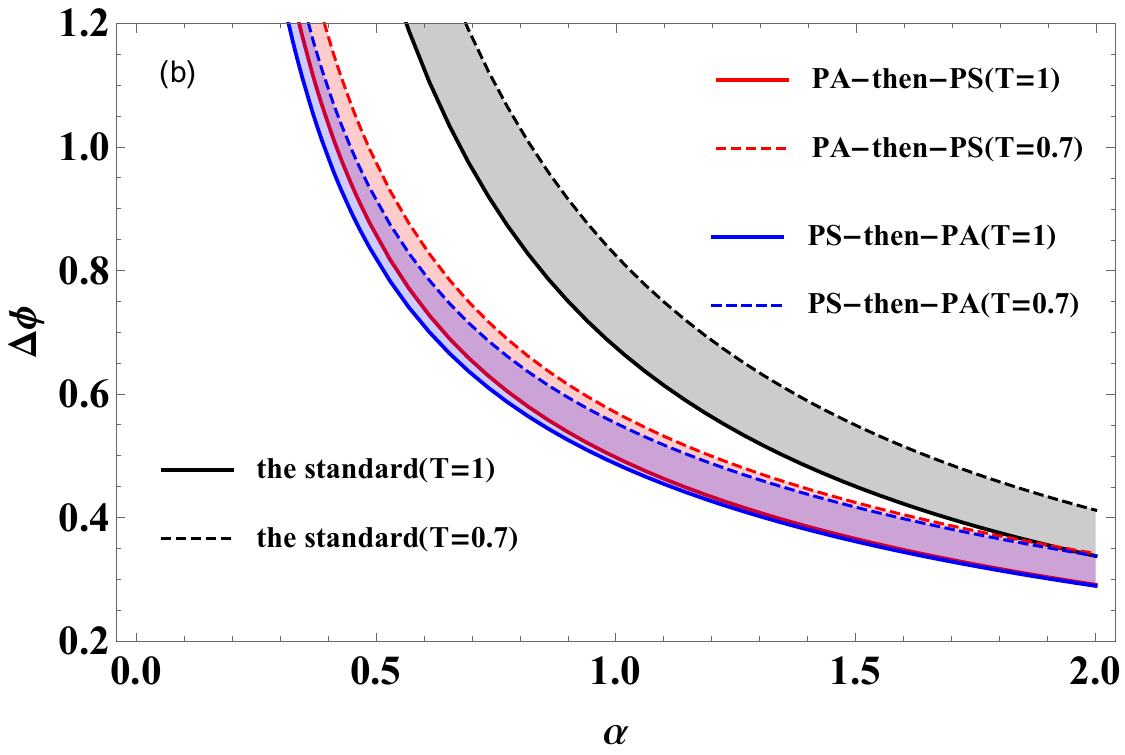}
\end{minipage}}
\caption{The comparisons for robustness against photon losses. (a) The phase
sensitivity as a function of $g$, with $\protect \alpha =1$ and $\protect \phi %
=0.6$. (b) The phase sensitivity as a function of $\protect \alpha $ , with $%
g=1$ and $\protect \phi =0.6$.}
\end{figure}

\subsection{Comparison with SQL and HL}

Additionally, we compare the phase sensitivity with SQL and HL in this
subsection. The SQL and HL are defined as $\Delta \phi _{SQL}=1/\sqrt{N_{j}}$
and $\Delta \phi _{HL}=1/N_{j}$, respectively. Here $N_{j}$ represents the
total mean photon number inside the interferometer before the second OPA for
each scheme \cite{e1,e3}, $j=1$ or $2$. $N_{j}$ can be calculated as%
\begin{eqnarray}
N_{1} &=&A_{1}^{2}\langle \psi _{in}|U_{S_{1}}^{\dagger }U_{B}^{\dagger
}U_{P_{1}}^{\dagger }\left( a^{\dagger }a+b^{\dagger }b\right)
U_{P_{1}}U_{B}U_{S_{1}}|\psi _{in}\rangle  \notag \\
&=&A_{1}^{2}(P_{3,3,0,0}+5P_{2,2,0,0}+4P_{1,1,0,0}  \notag \\
&&+P_{2,2,1,1}+3P_{1,1,1,1}+P_{0,0,1,1}),  \label{eq13}
\end{eqnarray}%
for the PA-then-PS and
\begin{eqnarray}
N_{2} &=&A_{2}^{2}\langle \psi _{in}|U_{S_{1}}^{\dagger }U_{B}^{\dagger
}U_{P_{2}}^{\dagger }\left( a^{\dagger }a+b^{\dagger }b\right)
U_{P_{2}}U_{B}U_{S_{1}}|\psi _{in}\rangle  \notag \\
&=&A_{2}^{2}(P_{3,3,0,0}+3P_{2,2,0,0}+P_{1,1,0,0}  \notag \\
&&+P_{2,2,1,1}+P_{1,1,1,1}),  \label{eq14}
\end{eqnarray}%
for the PS-then-PA, respectively.

For these two schemes at fixed $g$ and $\alpha ,$ we plot the phase
sensitivity $\Delta \phi $ as a function of $\phi $ for a comparison with
the SQL and the HL of the standard SU(1,1) interferometer. Our findings
demonstrate that (i) the original interferometer (without NCO) cannot
surpass the SQL. (ii) The NCO schemes are capable of surpassing the SQL
within a wide range, even in the presence of significant photon losses (Fig.
7(b)). This suggests that the NCO schemes show better robustness against
internal photon losses. (iii) The phase sensitivity of PS-then-PA is better
than that of PA-then-PS.

\begin{figure}[tph]
\label{Figure7} \centering%
\subfigure{
\begin{minipage}[b]{0.5\textwidth}
\includegraphics[width=0.83\textwidth]{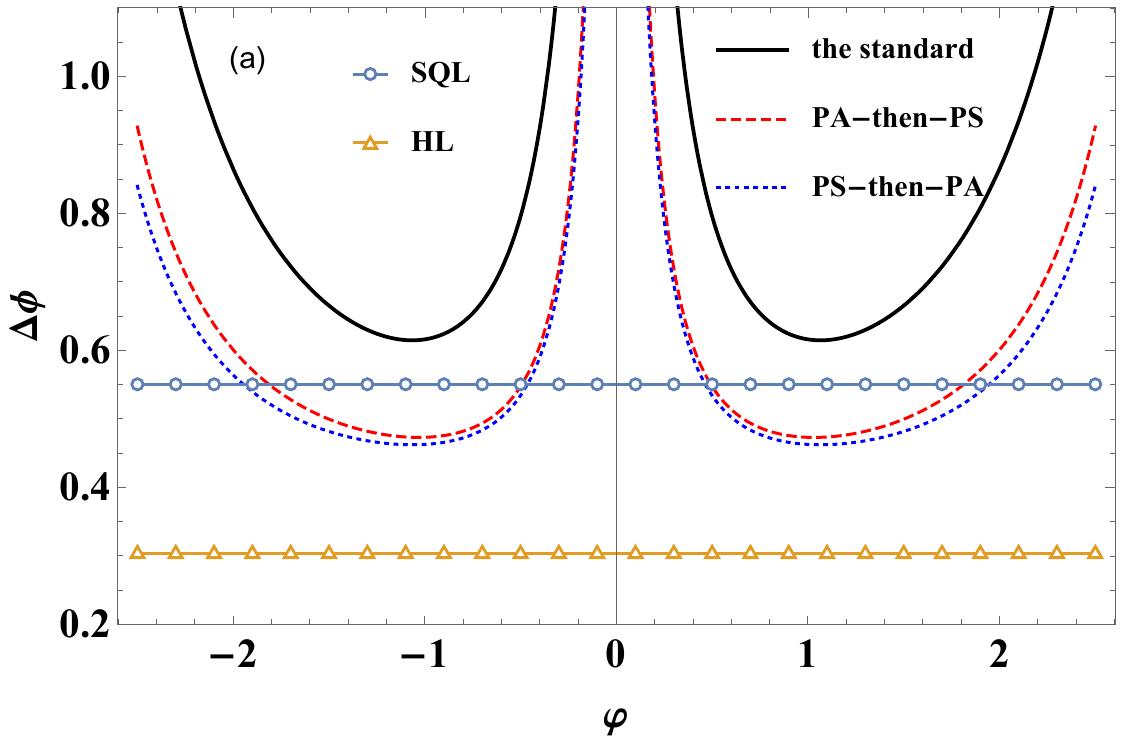}\\
\includegraphics[width=0.83\textwidth]{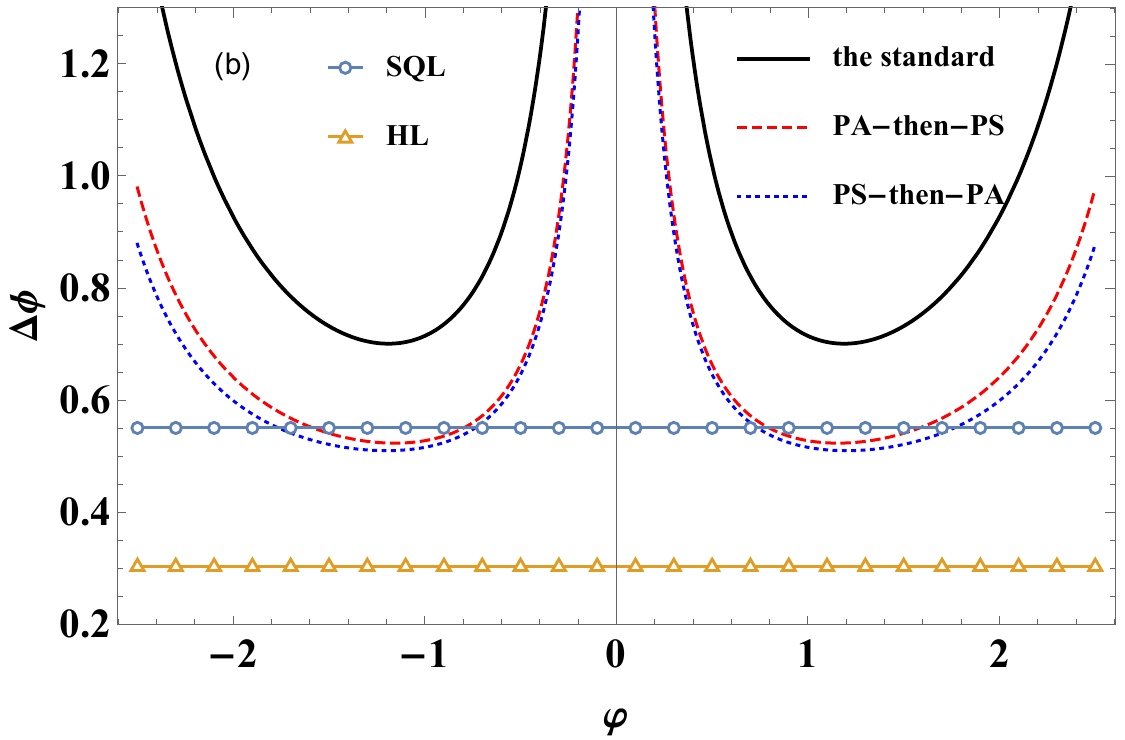}
\end{minipage}}
\caption{Comparison of phase sensitivity with the SQL and HL for fixed $%
g=0.7 $ and $\protect \alpha =1$. The blue circle is the SQL and the yellow
triangle is the HL. (a) $T=1$, (b) $T=0.7$.}
\end{figure}

\section{The quantum Fisher information}

In the previous discussions, we have explored the influence of NCO schemes
on phase sensitivity and the correlation between phase sensitivity and
relevant parameters using homodyne detection. It is crucial to recognize
that the discussed phase sensitivity is influenced by the chosen measurement
method. Hence, the question arises: how can we achieve maximum phase
sensitivity in an interferometer that is independent of the measurement
method used? This section shifts our focus to the QFI, which represents the
maximum information extracted from the interferometer system, regardless of
the measurement method employed. We will examine the QFI in ideal and
realistic scenarios, respectively.

\subsection{Ideal case}

For a pure state system, the QFI can be derived by \cite{e4}%
\begin{equation}
F_{j}=4\left[ \left \langle \Psi _{j}^{\prime }|\Psi _{j}^{\prime }\right
\rangle -\left \vert \left \langle \Psi _{j}^{\prime }|\Psi _{j}\right
\rangle \right \vert ^{2}\right] ,  \label{eq15}
\end{equation}%
where\ $\left \vert \Psi _{j}\right \rangle $ is the quantum state after
phase shift and before the second OPA, and $\left \vert \Psi _{j}^{\prime
}\right \rangle =\partial \left \vert \Psi _{j}\right \rangle /\partial \phi
.$ Then the QFI can be reformed as \cite{e4}
\begin{equation}
F_{j}=4\left \langle \Delta ^{2}n_{a}\right \rangle ,  \label{eq16}
\end{equation}%
where $\left \langle \Delta ^{2}n_{a}\right \rangle =\left \langle \Psi
_{j}\right \vert (a^{\dagger }a)^{2}|\Psi _{j}\rangle -(\left \langle \Psi
_{j}\right \vert a^{\dagger }a|\Psi _{j}\rangle )^{2}$.

\bigskip In the ideal NCO, the quantum state is given by $\left \vert \Psi
_{j}\right \rangle =A_{j}U_{\phi }U_{p_{j}}U_{S_{1}}\left \vert \alpha
\right \rangle _{a}\left \vert 0\right \rangle _{b}$, with $%
U_{P_{1}}=aa^{\dagger }$ ( $j=1$), $U_{P_{2}}=a^{\dagger }a$ ( $j=2$). Thus,
the QFI is derived as
\begin{eqnarray}
F_{1} &=&4\{A_{1}^{2}\left(
P_{4,4,0,0}+8P_{3,3,0,0}+14P_{2,2,0,0}+4P_{1,1,0,0}\right)  \notag \\
&&-\left[ A_{1}^{2}\left( P_{3,3,0,0}+5P_{2,2,0,0}+4P_{1,1,0,0}\right) %
\right] ^{2}\},  \label{eq17}
\end{eqnarray}%
for the PA-then-PS and
\begin{eqnarray}
F_{2} &=&4\{A_{2}^{2}\left(
P_{4,4,0,0}+6P_{3,3,0,0}+7P_{2,2,0,0}+P_{1,1,0,0}\right)  \notag \\
&&-\left[ A_{2}^{2}\left( P_{3,3,0,0}+3P_{2,2,0,0}+P_{1,1,0,0}\right) \right]
^{2}\},  \label{eq18}
\end{eqnarray}%
for the PS-then-PA, respectively. In the above equations, $T_{k}=1$. It is
possible to explore the connection between the QFI and the related
parameters using Eqs. (\ref{eq17}) and (\ref{eq18}).

Fig. 8 illustrates the QFI as a function of $g$ ($\alpha $) for a specific $%
\alpha $ ($g$). It is evident that a higher value of $g$ ($\alpha $)
corresponds to a greater QFI. Both PA-then-PS and PS-then-PA result in an
enhanced QFI due to the non-Gaussian nature. The QFI of PA-then-PS is
slightly higher than that of PS-then-PA in both figures. Moreover, we
observe that the improvement of QFI due to non-Gaussian operations increases
with the increase of the value $g$ (as shown in Fig. 8(a)), while it does
not significantly change with the variation of the value $\alpha $ (as shown
in Fig. 8(b)).

\begin{figure}[tph]
\label{Figure8} \centering%
\subfigure{
\begin{minipage}[b]{0.5\textwidth}
\includegraphics[width=0.83\textwidth]{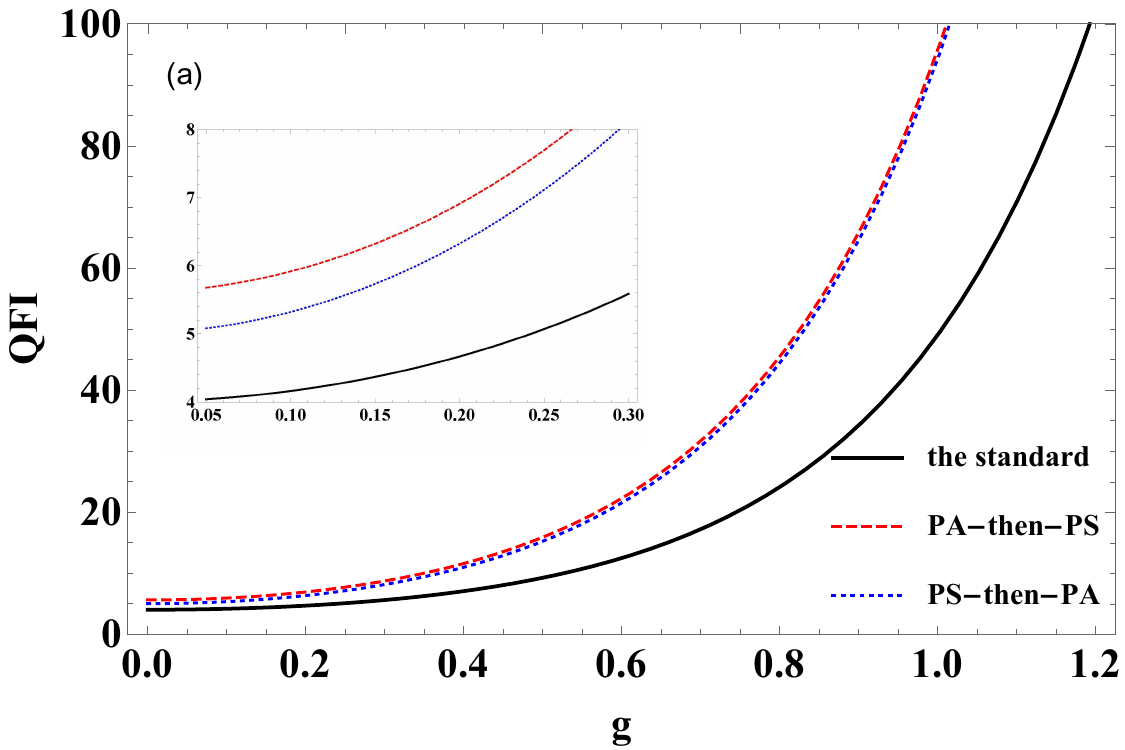}\\
\includegraphics[width=0.83\textwidth]{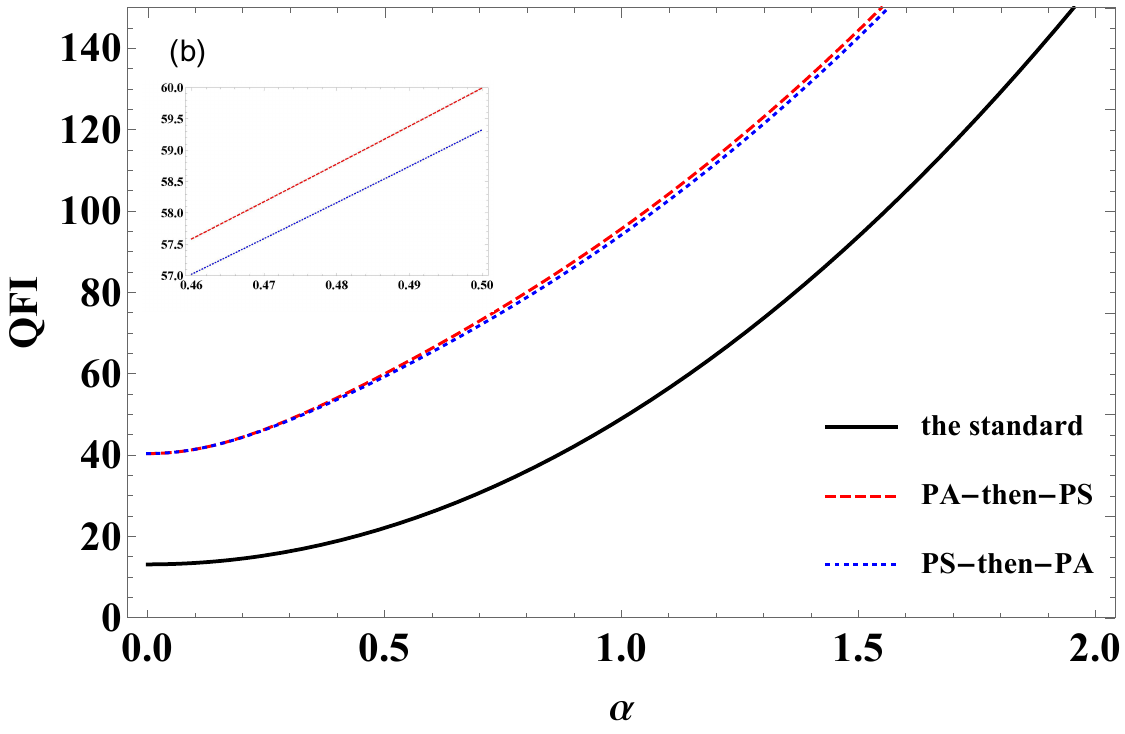}
\end{minipage}}
\caption{(a) The QFI as a function of $g$, with $\protect \alpha =1$. (b) The
QFI as a function of $\protect \alpha $, with $g=1$. }
\end{figure}

Actually, the QFI can be associated with the phase sensitivity through \cite%
{e5}%
\begin{equation}
\Delta \phi _{QCRB}=\frac{1}{\sqrt{vF}},  \label{eq19}
\end{equation}%
where $v$ represents the number of measurements. For simplicity, we set $v=1$%
. Another quantum limit, the QCRB \cite{c14,c15}, denoted as $\Delta \phi
_{QCRB}$, defines the ultimate limit for a set of probabilities derived from
measurements on a quantum system. It is an estimator implemented
asymptotically by a maximum likelihood estimator and provides a
detection-independent phase sensitivity. In order to better help us
understand how optimal the phase sensitivity obtained from the SU(1,1)
interferometer with the NCO really is, we compare the phase sensitivity $%
\Delta \phi $\ obtained by using the second OPA and homodyne detection with
the sensitivity $\Delta \phi _{QCRB}$\ obtained from the QFI. Fig. 9
illustrates the variation of $\Delta \phi _{QCRB}$ as a function of $g$ ($%
\alpha $) for a specific $\alpha $ ($g$). It is shown that $\Delta \phi
_{QCRB}$ improves with increasing $g$ and $\alpha $. Similarly, due to the
non-Gaussian nature, both PA-then-PS and PS-then-PA are able to improve $%
\Delta \phi _{QCRB}$. Furthermore, the improvement in $\Delta \phi _{QCRB}$
is more obvious for small coherent amplitude $\alpha $ (refer to Fig. 9(b)).
It is shown that for a smaller gain factor or a greater coherent amplitude,
the measurement-based sensitivity better reflects Fisher information
situation (described via $\Delta \phi _{QCRB}$).
\begin{figure}[tph]
\label{Figure9} \centering%
\subfigure{
\begin{minipage}[b]{0.5\textwidth}
\includegraphics[width=0.83\textwidth]{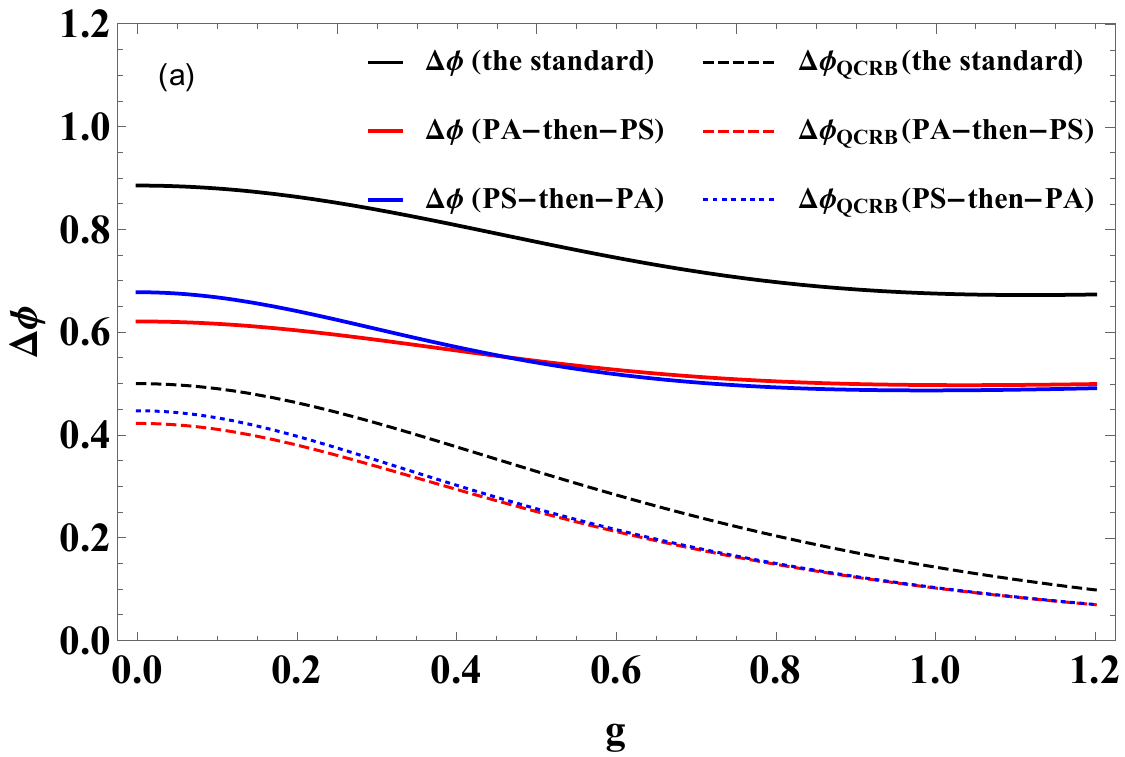}\\
\includegraphics[width=0.83\textwidth]{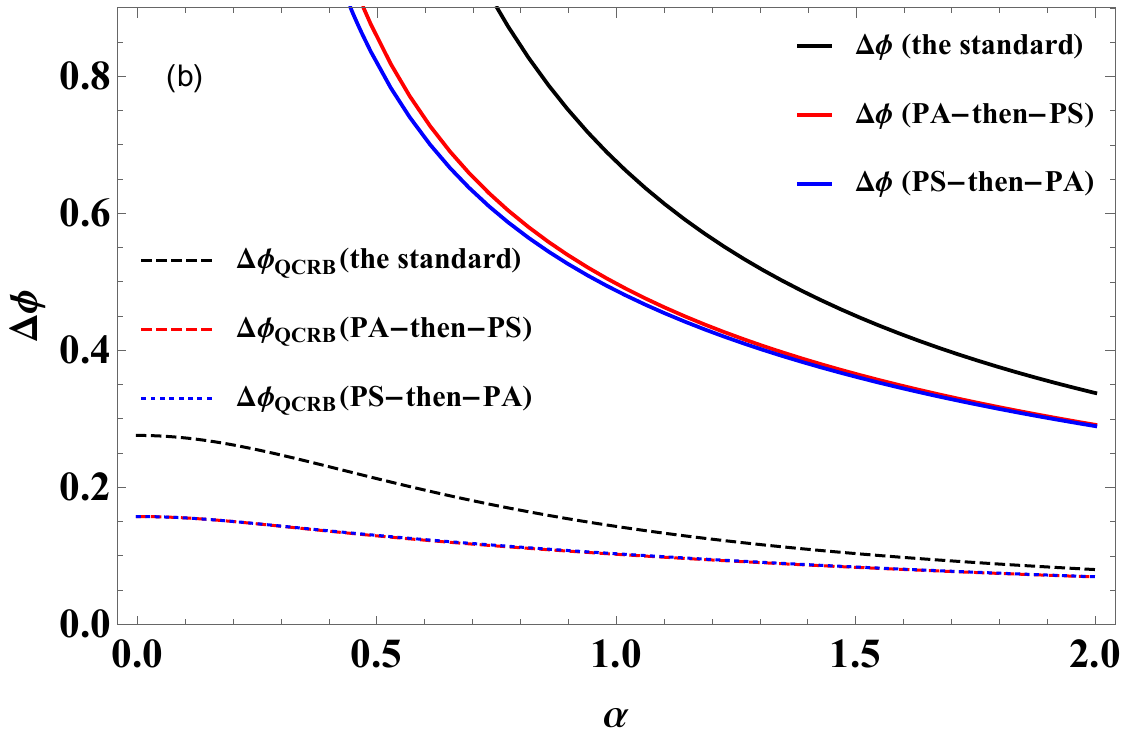}
\end{minipage}}
\caption{Comparisons of the phase sensitivity $\Delta \protect \phi $
obtained by homodyne detection with the ultimate sensitivity $\Delta \protect%
\phi _{QCRB}$ obtained from QFI.}
\end{figure}

\subsection{Photon losses case}

In this subsection, we extend our analysis to cover the QFI in the presence
of photon losses. Specifically, we examine homodyne detection on mode $a$,
which is susceptible to photon losses. Consequently, our attention is
directed toward the QFI of the system with photon losses in mode $a$, as
depicted in Fig. 10. Here, we should emphasize that the Fisher information
is obtained using the state preceding the second OPA, i.e., despite Fig. 10
featuring an SU(1,1) interferometer diagram, the second OPA is not
essential. For realistic quantum systems, we have demonstrated the
feasibility of computing the QFI with internal non-Gaussian operations
according to the method proposed by Escher \textit{et al}. \cite{e4}. Please
see Appendix B for the detailed process. The method is briefly summarized as
follows.

\begin{figure}[tph]
\label{Figure10} \centering \includegraphics[width=0.9%
\columnwidth]{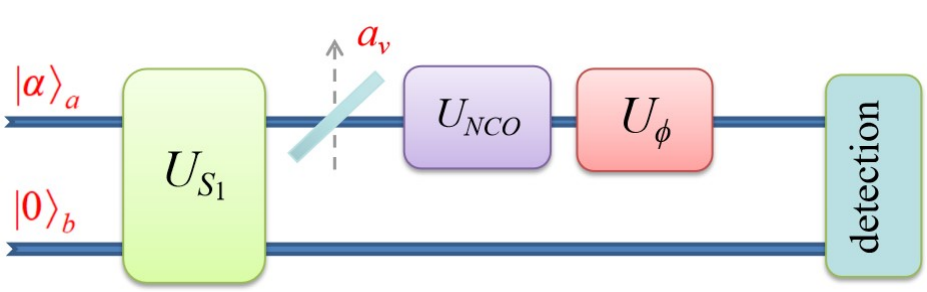} \
\caption{Schematic diagram of the photon losses on mode $a$. The losses
occurs before the NCO.}
\end{figure}

For the case of photon losses, we can treat the system as a pure state in an
extended space, similar to Eq. (\ref{eq3}). Then following Eq. (\ref{eq15}),
we can obtain the QFI under the pure state, denoted as $C_{Q_{j}}$, which is
larger or equal to the QFI $F_{L_{j}}$ for mixed state (our consideration),
i.e., $F_{L_{j}}\leq C_{Q_{j}}$. $C_{Q_{j}}$ is the QFI before optimizing
over all possible measurements, i.e.,%
\begin{equation}
C_{Q_{j}}=4\left[ \left \langle \psi \right \vert \hat{H}_{1_{j}}\left \vert
\psi \right \rangle \right. -\left. \left \vert \left \langle \psi \right
\vert \hat{H}_{2_{j}}\left \vert \psi \right \rangle \right \vert ^{2}\right]
,  \label{eq21}
\end{equation}%
where $\hat{H}_{1_{j}}$ and $\hat{H}_{2_{j}}$ are defined as%
\begin{eqnarray}
\hat{H}_{1_{j}} &=&B_{j}^{2}\overset{\infty }{\underset{l=0}{\sum }}\frac{d}{%
d\phi }\Pi _{l}^{\dagger }\left( \eta ,\phi ,\lambda \right)
U_{p_{j}}^{\dagger }U_{p_{j}}\frac{d}{d\phi }\Pi _{l}\left( \eta ,\phi
,\lambda \right) ,  \label{22} \\
\hat{H}_{2_{j}} &=&iB_{j}^{2}\overset{\infty }{\underset{l=0}{\sum }}\left[
\frac{d}{d\phi }\Pi _{l}^{\dagger }\left( \eta ,\phi ,\lambda \right) \right]
U_{p_{j}}^{\dagger }U_{p_{j}}\Pi _{l}\left( \eta ,\phi ,\lambda \right) .
\label{23}
\end{eqnarray}%
Here $B_{j}$ are normalization factors shown in Eq. (B10), and $\Pi
_{l}\left( \eta ,\phi ,\lambda \right) $ is the phase-dependent Krause
operator shown in Eq. (B8), satisfying $\sum \Pi _{l}^{\dagger }\left( \eta
,\phi ,\lambda \right) \Pi _{l}\left( \eta ,\phi ,\lambda \right) =1$, with $%
\lambda =0$ and $\lambda =-1$ representing the photon losses before the
phase shifter and after the phase shifter, respectively. $\eta $ is related
to the dissipation factor with $\eta =1$ and $\eta =0$ being the cases of
complete lossless and absorption, respectively. Partiularly, Eqs. (\ref{eq21}%
), (\ref{22}) and (\ref{23}) just reduce to these in Ref. \cite{e4}, when
there is no non-Gaussian operations. Following the spirit of Ref. \cite{e4},
we can further obtain the minimum value of $C_{Q_{j}}$ by optimizing over $%
\lambda $, corresponding to $F_{L_{j}}$, i.e., $F_{L_{j}}=\min_{\Pi
_{l}\left( \eta ,\phi ,\lambda \right) }C_{Q_{j}}\leq C_{Q_{j}}$. See
Appendix B about more details.

Next, we further analyze the effects of each parameter on the QFI of the NCO
schemes under photon losses by numerical calculation. Fig. 11 is plotted the
QIF and QCRB as a function of transmittance $\eta ,$ from which it is
observed that the QFI increases with the rising transmittance $\eta $, and
the NCO can enhance the QFI. This increase can be attributed to the NCO
raising the number of photons internally, resulting in higher quantum
information, similar to the ideal case. For both non-Gaussian operations,
the improved QFI increases with the transmittance $\eta $. It is interesting
that, over a wide range of about $0<$ $\eta $ $<0.85$, the PS-then-PA
exhibits a bigger QFI or higher precision than the PA-then-PS. However, as $%
\eta $ approaches $1$, the PA-then-PS demonstrates a superior QFI/QCRB
within the range of about $0.85<$ $\eta $ $<1$. This implies that the
PS-then-PA presents better performance than PA-then-PS under high
dissipation situation.

\begin{figure}[tph]
\label{Figure11} \centering%
\subfigure{
\begin{minipage}[b]{0.5\textwidth}
\includegraphics[width=0.83\textwidth]{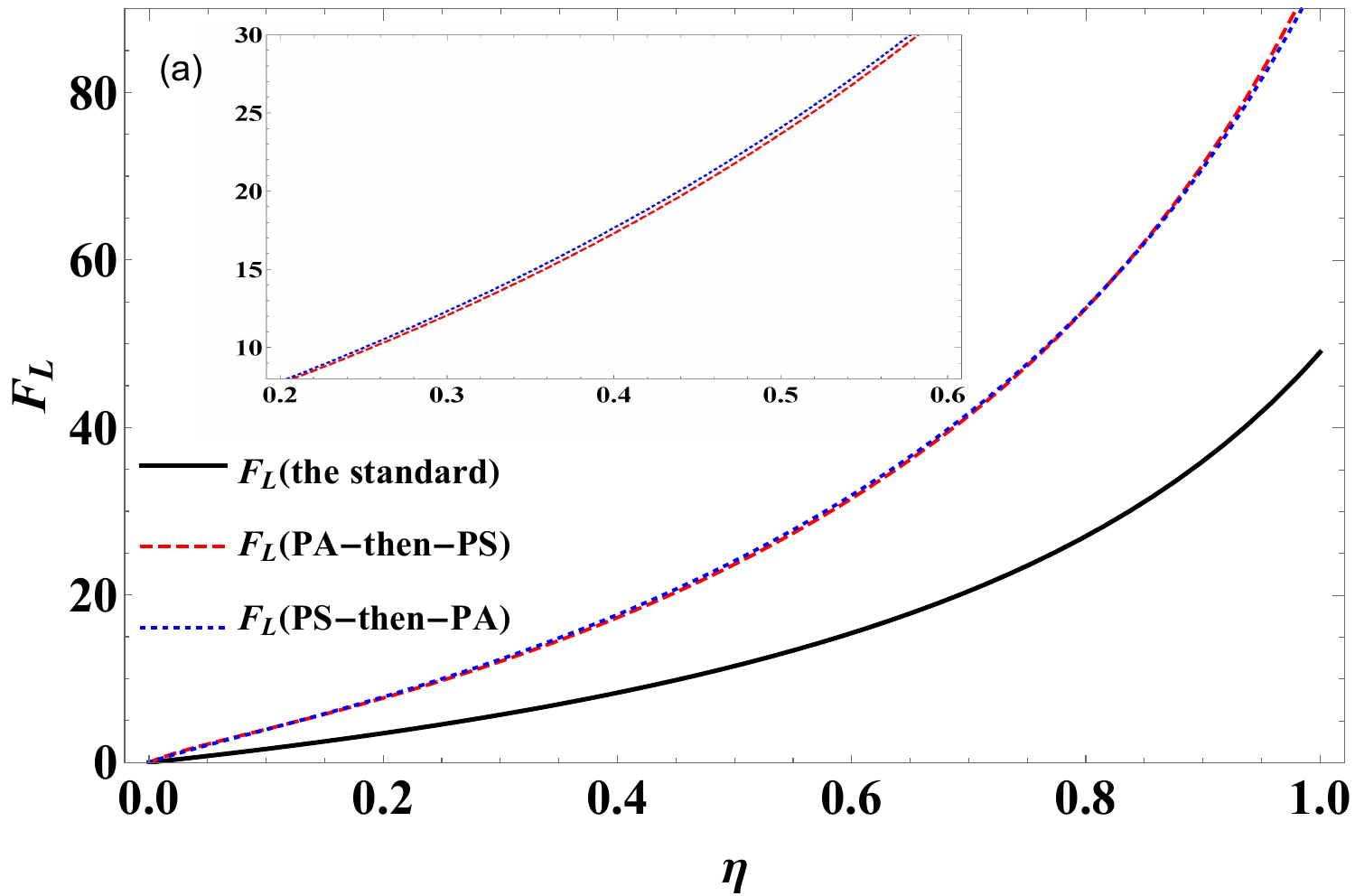}\\
\includegraphics[width=0.83\textwidth]{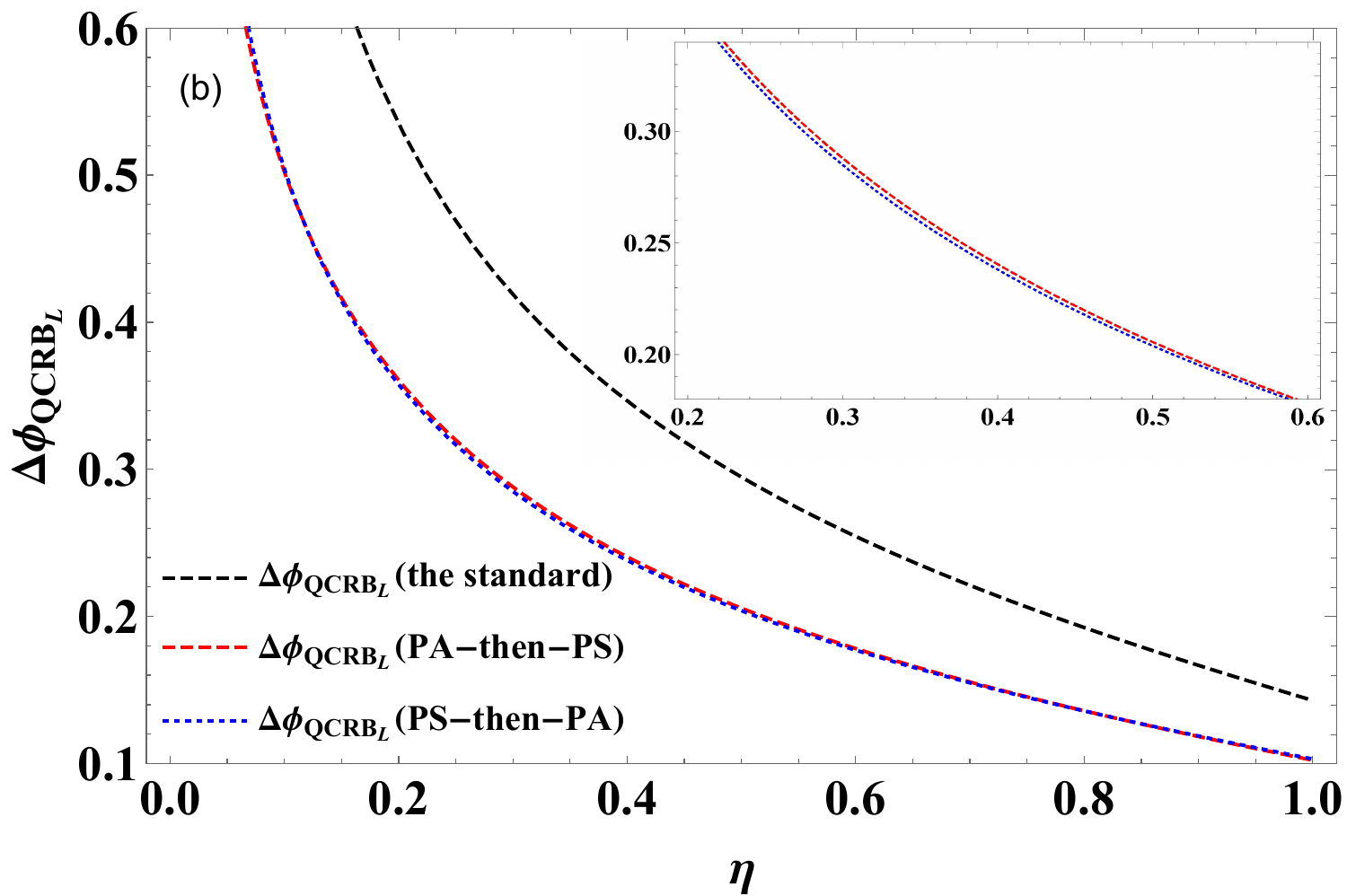}
\end{minipage}}
\caption{The $F_{L}$ and $\Delta \protect \phi _{QCRB_{L}}$ as functions of
transmittance $\protect \eta $, with $g=1$ and $\protect \alpha =1$.}
\end{figure}

To explore the underlying reasons for the above case, we further examine the
non-classicality of the NCO by the negative volume of Wigner Function (WF)
\cite{e31}. For simplicity, we only consider the WF of ideal quantum states
after non-Gaussian operations. Some details are summarized in Appendix C
about the WF. Fig. 12 illustrates the WF in phase space corresponding to two
different operations. It is clear that, (i) both non-Gaussian operations can
increase the negative volume of WF, i.e., increase the non-classicality \cite%
{e31}. (ii) For given $\alpha $\ and $g$, the PS-then-PA presents a much
bigger negative volume than the PA-then-PS. For example, for $\alpha =1$\
and $g=0.6,0.8,1.0,1.2$, the negative volumes are 0.034/0.009, 0.033/0.014,
0.031/0.017, 0.030/0.020 for PS-then-PA/PA-then-PS, respectively. These
observations suggest that the non-Gaussian operation increases the
non-classicality, and the stronger the nonclassicality of the internal
non-Gaussian operation, the more effective it is in suppressing the effect
of the internal high noise.

\begin{figure}[tph]
\label{Figure12} \centering%
\subfigure{
\begin{minipage}[b]{0.45\textwidth}
\includegraphics[width=0.27\textwidth]{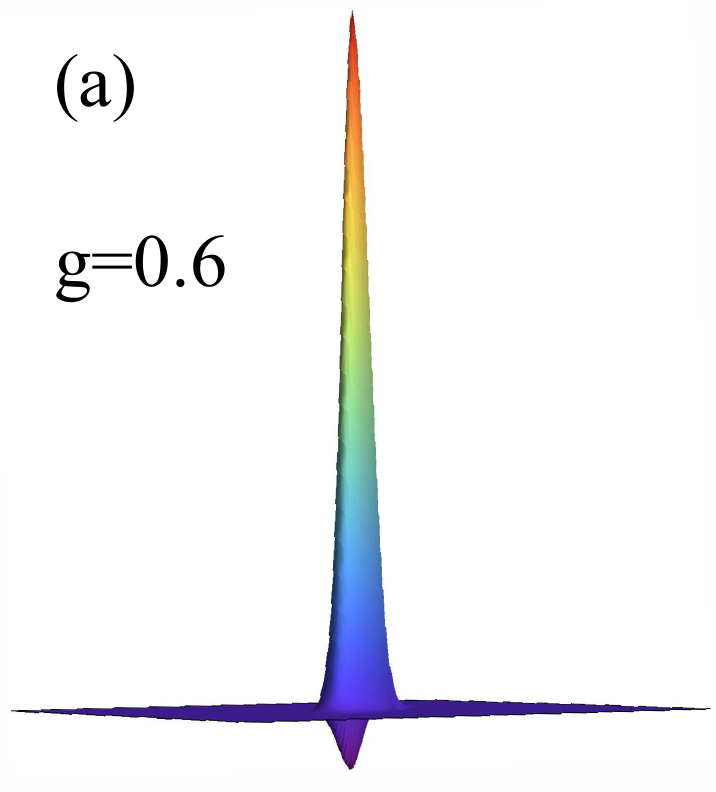}\includegraphics[width=0.27\textwidth]{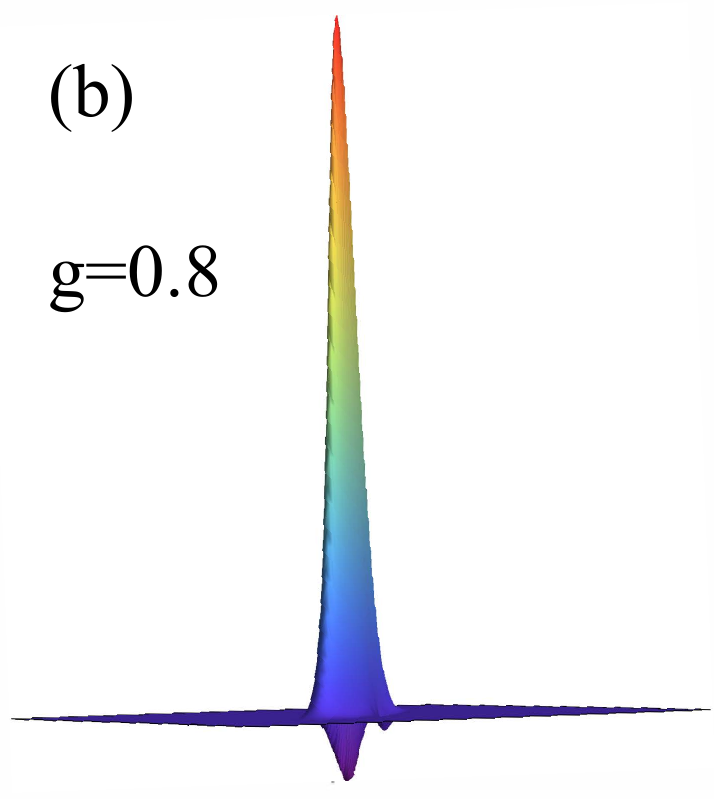}\includegraphics[width=0.27\textwidth]{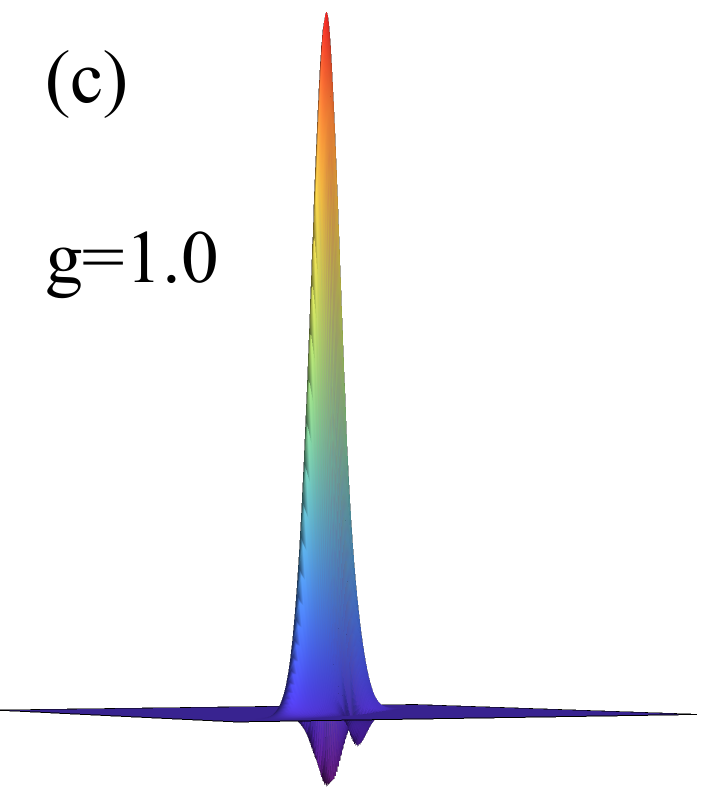}\includegraphics[width=0.27\textwidth]{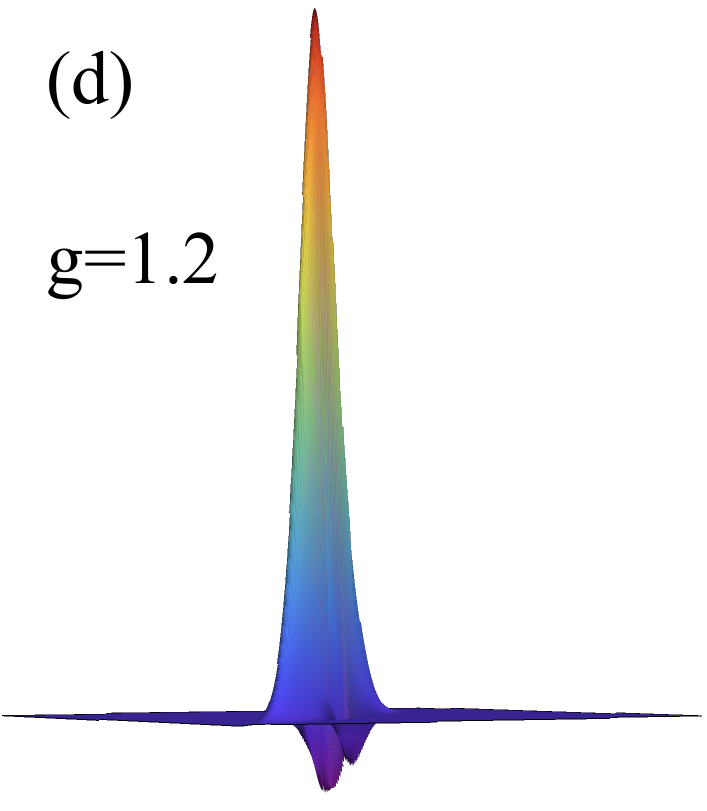}\\
\includegraphics[width=0.27\textwidth]{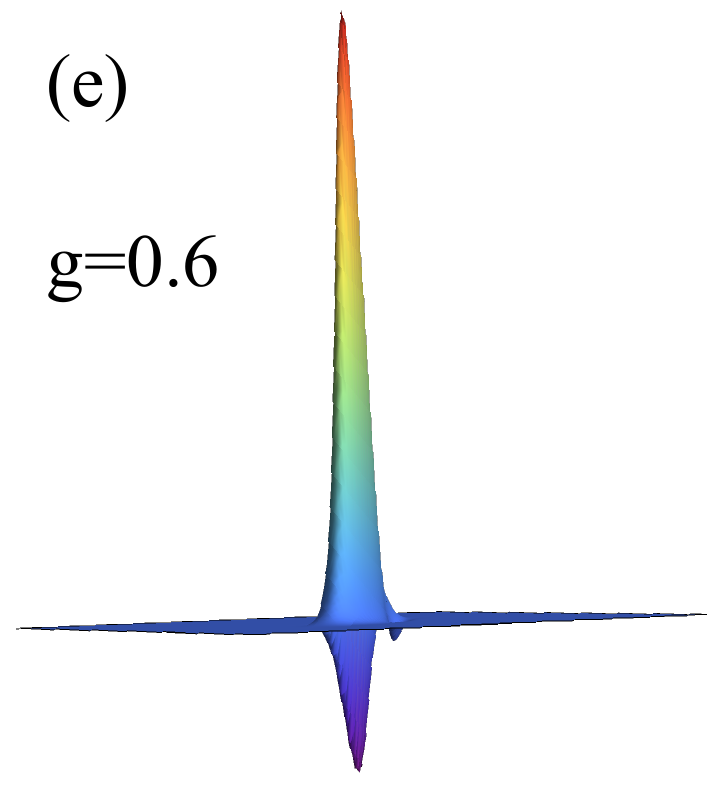}\includegraphics[width=0.27\textwidth]{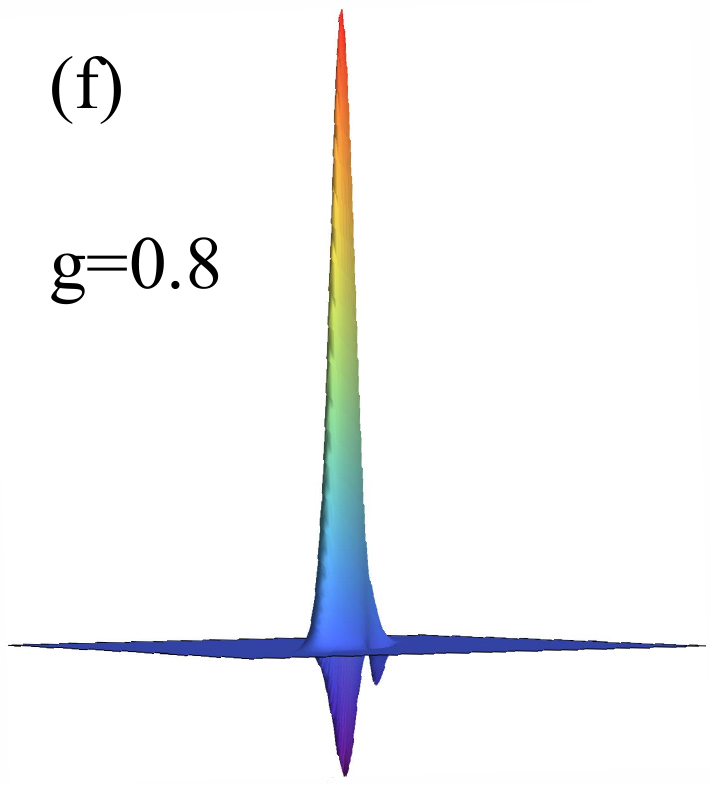}\includegraphics[width=0.27\textwidth]{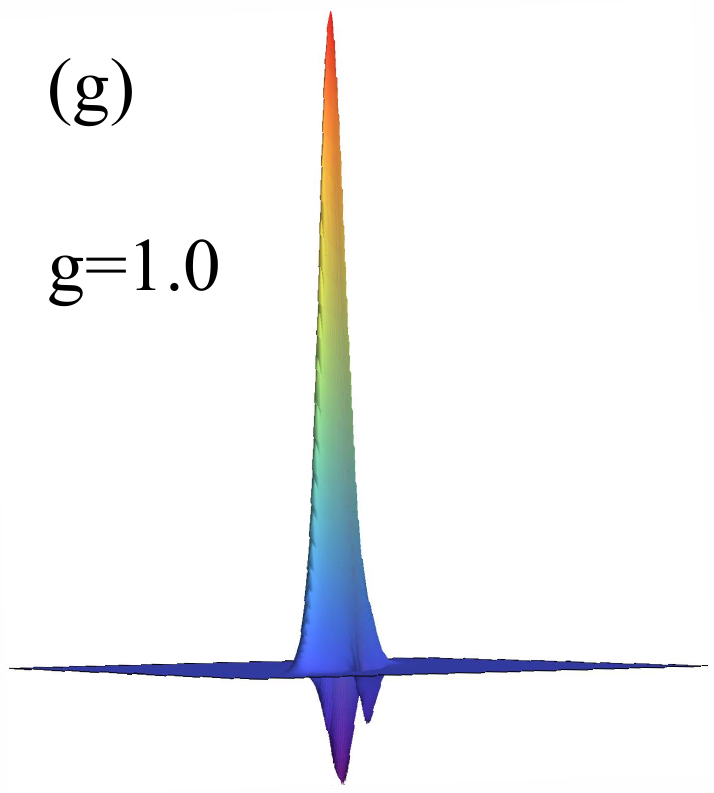}\includegraphics[width=0.27\textwidth]{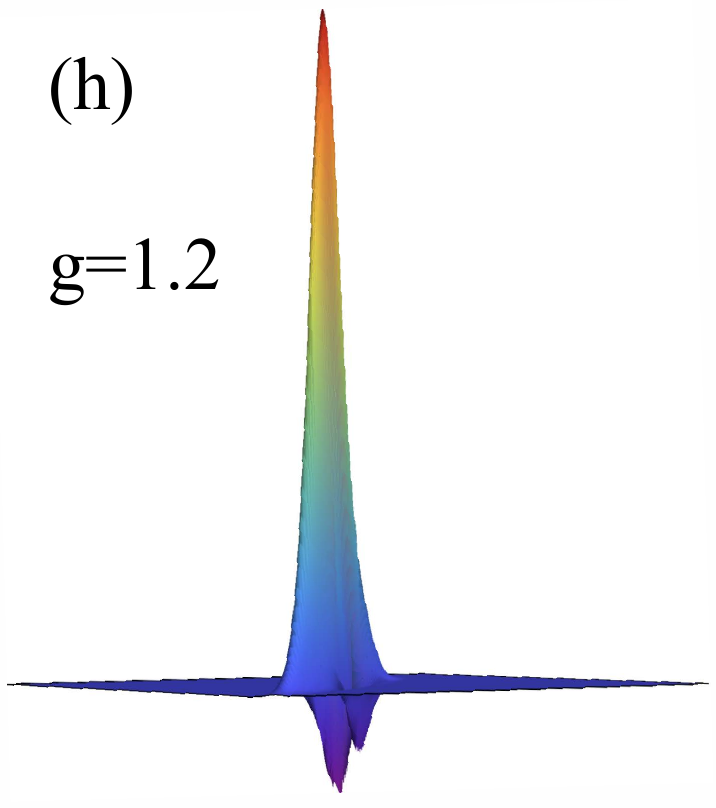}
\end{minipage}}
\caption{The WF in phase space for quantum states after the NCO with $%
\protect \alpha =1$. (a)-(d) for the PA-then-PS and (e)-(h) for the
PS-then-PA, with several different $g=0.6,0.8,1.0,1.2$ (from left to right).}
\end{figure}

Similar to the ideal case, Fig. 13 illustrates the QFI as a function of $g$ (%
$\alpha $) for a given $\alpha $ ($g$), under the loss case with $\eta =0.6$%
. Some similar results to Fig. 8 can be obtained, not shown here. Different
from the ideal case in Fig. 8, the PS-then-PA scheme performs better than
PA-then-PS when $g$ is larger, shown in Fig. 13(a). This case is also true
for the QFI with $\alpha $, shown in Fig. 13(b). These two cases are almost
the opposite to the previous ideal situation. The reason may be that the
PS-then-PA operation prepares the higher non-classical states, which are
more conducive to improve the measurement accuracy, especially in the
presence of high photon losses.

\begin{figure}[tph]
\label{Figure13} \centering%
\subfigure{
\begin{minipage}[b]{0.5\textwidth}
\includegraphics[width=0.83\textwidth]{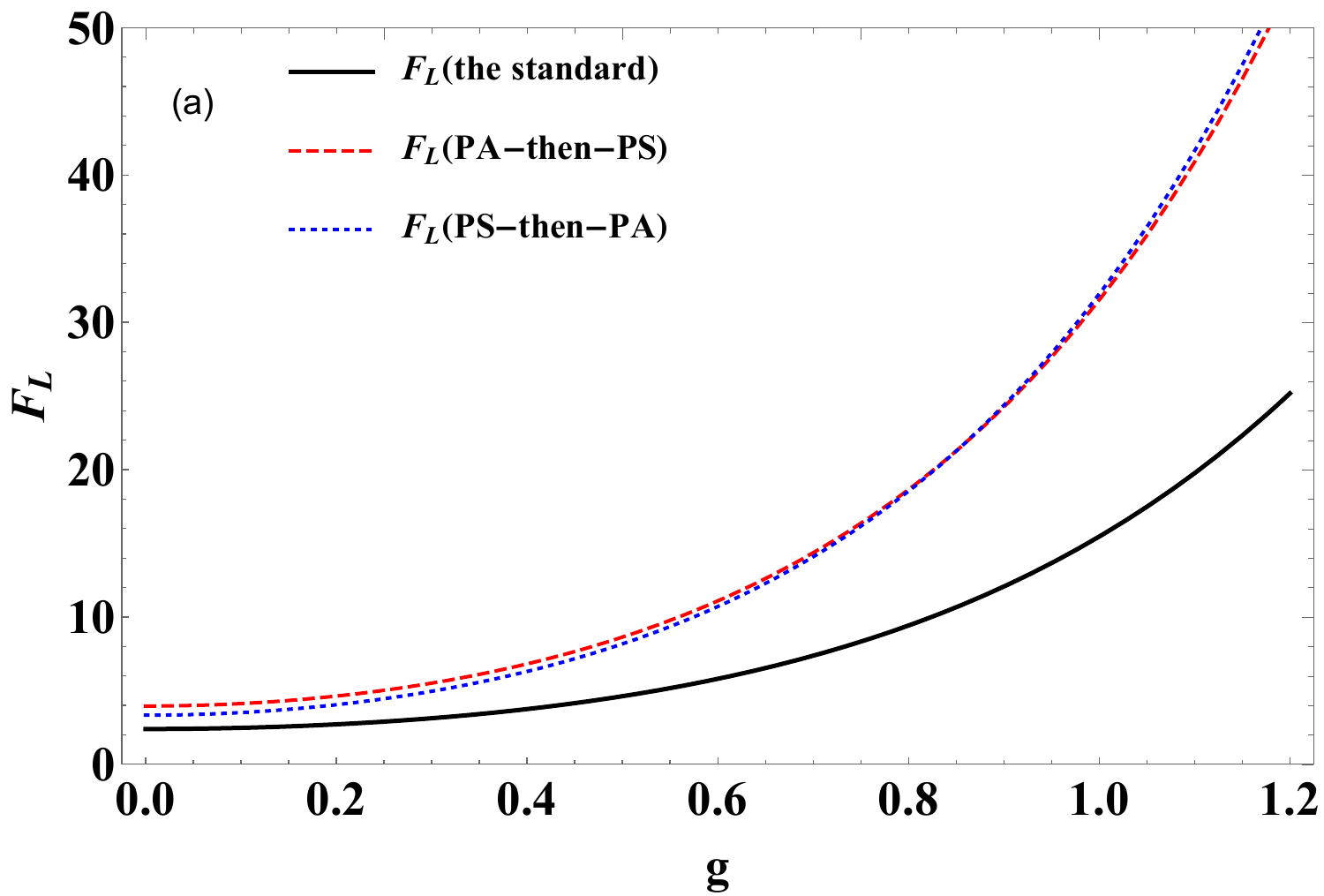}\\
\includegraphics[width=0.83\textwidth]{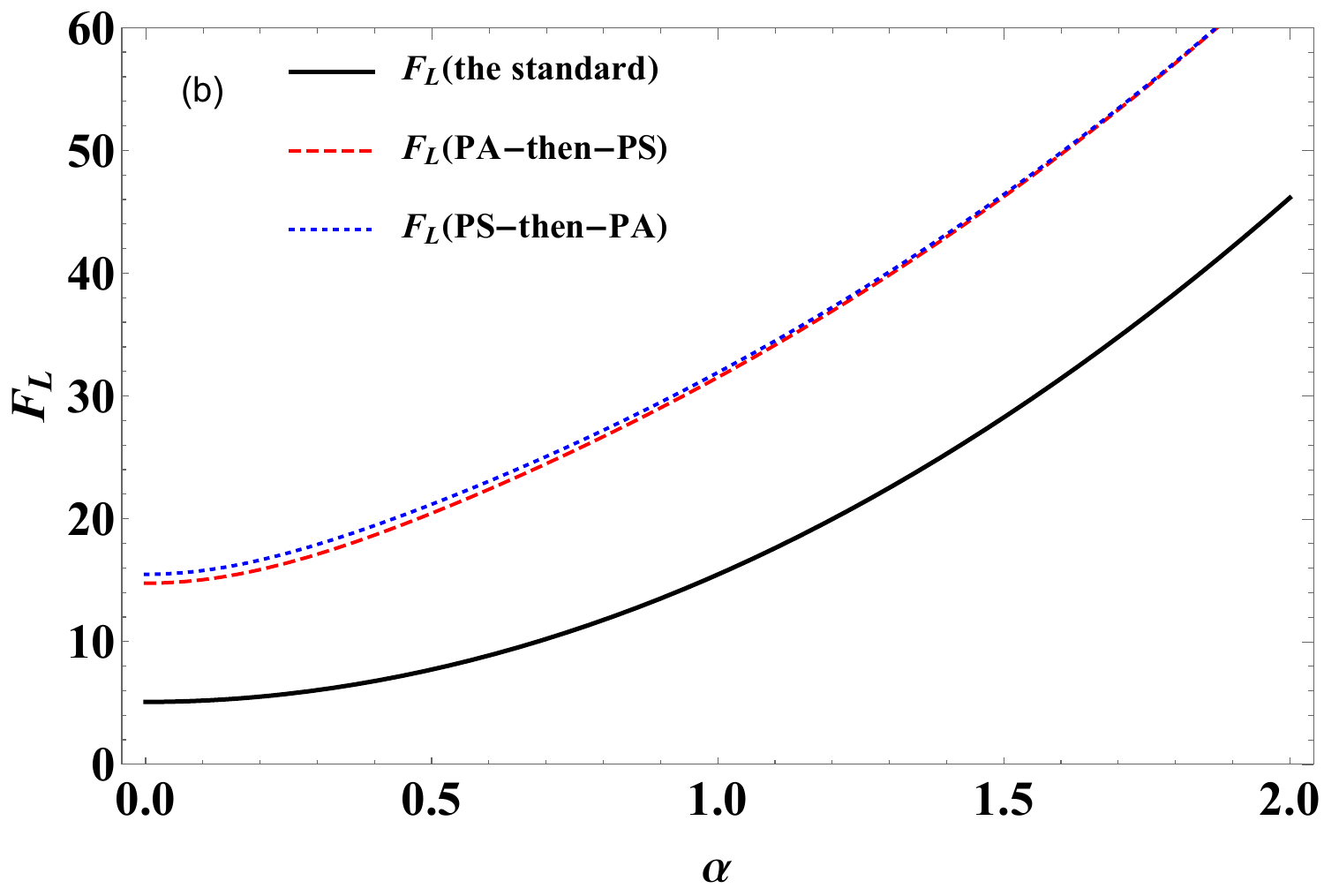}
\end{minipage}}
\caption{(a) The $F_{L}$ as a function of $g$, with $\protect \alpha =1$ \
and $\protect \eta =0.6$. (b) The $F_{L}$ as a function of $\protect \alpha $,
with $g=1$ \ and $\protect \eta =0.6$.}
\end{figure}

\section{Conclusion}

In this paper, we have analyzed the effects of NCO schemes on the phase
sensitivity, the QFI and the QCRB in both ideal and photon losses cases.
Additionally, we have investigated the effects of the gain coefficient $g$
of OPA, the coherent state amplitude $\alpha $ and the transmittance $T_{k}$
of BS on the performance of the system. Through analytical comparison, we
have verified that the NCO schemes can improve the measurement accuracy of
the SU(1,1) interferometer and enhance the robustness against internal
photon losses. The non-Gaussian operations can elevate the total mean photon
number of the SU(1,1) interferometer, consequently reinforcing intramode
correlations and quantum entanglement between the two modes.

We further analyze the differences between the two non-Gaussian operations.
Concerning the phase sensitivity, the improvement of PS-then-PA is superior
in both ideal and photon losses cases. In terms of the QFI and QCRB, in the
ideal case, the PA-then-PS is slightly outperforms the PS-then-PA. However,
in the photon losses case, the PS-then-PA demonstrates a greater advantage.

In summary, the NCO schemes play a role in overcoming the internal photon
losses within SU(1,1) interferometers and in improving the accuracy of
quantum measurements. This study highlights the potential of the
non-Gaussian operations as valuable tools for improving the performance of
quantum metrology and information processing systems. It should be noted
that we mainly pay attention to an ideal PS/PA case. Actually, there are
some methods to realize these operations. The different experimental
parameters will impact the performance, which will be further examined in
the near future.

\begin{acknowledgments}
This work is supported by the National Natural Science Foundation of China
(Grants No. 11964013 and No. 12104195) and Key project of Natural Science Foundation of Jiangxi Province,
Jiangxi Provincial Key Laboratory of Advanced Electronic Materials and Devices (No. 2024SSY03011).
\end{acknowledgments}\bigskip

\textbf{APPENDIX A : THE PHASE SENSITIVITY WITH NCO}\bigskip

In this Appendix, we give the calculation formulas of the phase sensitivity
with NCO as follows
\begin{equation}
\Delta \phi _{1}=\frac{\sqrt{\left \langle \Psi _{out}^{1}\right \vert
\left( a^{\dagger }+a\right) ^{2}\left \vert \Psi _{out}^{1}\right \rangle
-\left \langle \Psi _{out}^{1}\right \vert \left( a^{\dagger }+a\right)
\left \vert \Psi _{out}^{1}\right \rangle ^{2}}}{|\partial \left \langle
\Psi _{out}^{1}\right \vert \left( a^{\dagger }+a\right) \left \vert \Psi
_{out}^{1}\right \rangle /\partial \phi |}.  \tag{A1}
\end{equation}%
Here, the output state $\left \vert \Psi _{out}^{1}\right \rangle $ is given
by Eq. (\ref{eq3}), so the expectations related to the phase sensitivity in
PA-then-PS are specifically calculated as follows \cite{c13}
\begin{align}
& \left \langle \Psi _{out}^{1}\right \vert \left( a^{\dagger }+a\right)
\left \vert \Psi _{out}^{1}\right \rangle  \notag \\
=& A_{1}^{2}[e^{-i\phi }\cosh g\left(
P_{3,2,0,0}+4P_{2,1,0,0}+2P_{1,0,0,0}\right)  \notag \\
& +\sinh g\left( P_{2,2,0,1}+3P_{1,1,0,1}+P_{0,0,0,1}\right)  \notag \\
& +e^{i\phi }\cosh g\left( P_{2,3,0,0}+4P_{1,2,0,0}+2P_{0,1,0,0}\right)
\notag \\
& +\sinh g\left( P_{2,2,1,0}+3P_{1,1,1,0}+P_{0,0,1,0}\right) ],  \tag{A2}
\end{align}%
and
\begin{align}
& \left \langle \Psi _{out}^{1}\right \vert \left( a^{\dagger }+a\right)
^{2}\left \vert \Psi _{out}^{1}\right \rangle  \notag \\
=& A_{1}^{2}[e^{-2i\phi }\cosh ^{2}g\left(
P_{4,2,0,0}+5P_{3,1,0,0}+3P_{2,0,0,0}\right)  \notag \\
& e^{2i\phi }\cosh ^{2}g\left( P_{2,4,0,0}+5P_{1,3,0,0}+3P_{0,2,0,0}\right)
\notag \\
& +2\cosh ^{2}g\left( P_{3,3,0,0}+5P_{2,2,0,0}+4P_{1,1,0,0}\right)  \notag \\
& +2e^{-i\phi }\sinh g\cosh g(P_{3,2,0,1}+4P_{2,1,0,1}  \notag \\
& +2P_{1,0,0,1}+P_{3,2,1,0}+4P_{2,1,1,0}+2P_{1,0,1,0})  \notag \\
& +2e^{i\phi }\sinh g\cosh g(P_{2,3,1,0}+4P_{1,2,1,0}  \notag \\
& +2P_{0,1,1,0}+P_{2,3,0,1}+4P_{1,2,0,1}+2P_{0,1,0,1})  \notag \\
& +\sinh ^{2}g(P_{2,2,0,2}+3P_{1,1,0,2}+P_{0,0,0,2}  \notag \\
& +P_{2,2,2,0}+3P_{1,1,2,0}+P_{0,0,2,0}+2P_{2,2,1,1}  \notag \\
& +6P_{1,1,1,1}+2P_{0,0,1,1}+2P_{2,2,0,0}  \notag \\
& +6P_{1,1,0,0}+2)+A_{1}^{-2}].  \tag{A3}
\end{align}%
The phase sensitivity with the PS-then-PA can be calculated as
\begin{equation}
\Delta \phi _{2}=\frac{\sqrt{\left \langle \Psi _{out}^{2}\right \vert
\left( a^{\dagger }+a\right) ^{2}\left \vert \Psi _{out}^{2}\right \rangle
-\left \langle \Psi _{out}^{2}\right \vert \left( a^{\dagger }+a\right)
\left \vert \Psi _{out}^{2}\right \rangle ^{2}}}{|\partial \left \langle
\Psi _{out}^{2}\right \vert \left( a^{\dagger }+a\right) \left \vert \Psi
_{out}^{2}\right \rangle /\partial \phi |},  \tag{A4}
\end{equation}%
where the output state $\left \vert \Psi _{out}^{2}\right \rangle $ is given
by Eq. (\ref{eq4}), and the expectations associated with the phase
sensitivity for the PS-then-PA can similarly be calculated as follows
\begin{align}
& \left \langle \Psi _{out}^{2}\right \vert \left( a^{\dagger }+a\right)
\left \vert \Psi _{out}^{2}\right \rangle  \notag \\
=& A_{2}^{2}[e^{-i\phi }\cosh g\left( P_{3,2,0,0}+2P_{2,1,0,0}\right)  \notag
\\
& +\sinh g\left( P_{2,2,0,1}+P_{1,1,0,1}\right)  \notag \\
& +e^{i\phi }\cosh g\left( P_{2,3,0,0}+2P_{1,2,0,0}\right)  \notag \\
& +\sinh g\left( P_{2,2,1,0}+P_{1,1,1,0}\right) ],  \tag{A5}
\end{align}%
and
\begin{align}
& \left \langle \Psi _{out}^{2}\right \vert \left( a^{\dagger }+a\right)
^{2}\left \vert \Psi _{out}^{2}\right \rangle  \notag \\
=& A_{2}^{2}[e^{-2i\phi }\cosh ^{2}g\left( P_{4,2,0,0}+3P_{3,1,0,0}\right)
\notag \\
& +e^{2i\phi }\cosh ^{2}g\left( P_{2,4,0,0}+3P_{1,3,0,0}\right)  \notag \\
& +2\cosh ^{2}g\left( P_{3,3,0,0}+3P_{2,2,0,0}+P_{1,1,0,0}\right)  \notag \\
& +2e^{-i\phi }\sinh g\cosh g(P_{3,2,1,0}+2P_{2,1,1,0}  \notag \\
& +P_{3,2,0,1}+2P_{2,1,0,1})  \notag \\
& +2e^{i\phi }\sinh g\cosh g(P_{2,3,0,1}+2P_{1,2,0,1}  \notag \\
& +P_{2,3,1,0}+2P_{1,2,1,0})  \notag \\
& +\sinh ^{2}g(P_{2,2,2,0}+P_{1,1,2,0}+P_{2,2,0,2}  \notag \\
& +P_{1,1,0,2}+2P_{2,2,1,1}+2P_{1,1,1,1}  \notag \\
& +2P_{2,2,0,0}+2P_{1,1,0,0})+A_{2}^{-2}].  \tag{A6}
\end{align}%
\bigskip

\textbf{APPENDIX B : THE QFI WITH PHOTON LOSSES}\bigskip

Here, we further examine the QFI with photon losses for the system as shown
in Fig. 10. After the first OPA $U_{S_{1}}$, the photon losses, the
non-Gaussian operation $U_{P_{j}}$ ( $j=1$ or $2$), and before the
detection, the output state in an expanded space can be given by
\begin{equation}
\left \vert \Psi _{E_{j}}\right \rangle =B_{j}U_{\phi }U_{p_{j}}U_{B}\left
\vert 0\right \rangle _{a_{v}}\left \vert \psi \right \rangle ,  \tag{B1}
\end{equation}%
a form of pure state, where $\left \vert \psi \right \rangle
=U_{S_{1}}\left
\vert \alpha \right \rangle _{a}\left \vert 0\right \rangle
_{b}, $ and $B_{j}$ is the normalized factor, determined by Tr$\left \vert
\Psi _{E_{j}}\right \rangle \left \langle \Psi _{E_{j}}\right \vert =1$.

For a pure state system, the QFI can be calculated using Eq. (\ref{eq15}),
denoted as $C_{Q_{j}}$. Substituting Eq. (B1) into Eq. (\ref{eq15}) yields
\begin{equation}
C_{Q_{j}}=4\left[ \left \langle \psi \right \vert \hat{H}_{1_{j}}\left \vert
\psi \right \rangle \right. -\left. \left \vert \left \langle \psi \right
\vert \hat{H}_{2_{j}}\left \vert \psi \right \rangle \right \vert ^{2}\right]
,  \tag{B2}
\end{equation}%
where $\hat{H}_{1_{j}}$ and $\hat{H}_{2_{j}}$ are operators, defined as
\begin{align}
\hat{H}_{1_{j}}& =B_{j}^{2}\left. _{a_{v}}\left \langle 0\right \vert \frac{%
dU_{B}^{\dagger }U_{p_{j}}^{\dagger }U_{\phi }^{\dagger }}{d\phi }\frac{%
dU_{\phi }U_{p_{j}}U_{B}}{d\phi }\left \vert 0\right \rangle _{a_{v}}\right.
,  \tag{B3} \\
\hat{H}_{2_{j}}& =iB_{j}^{2}\left. _{a_{v}}\left \langle 0\right \vert \frac{%
dU_{B}^{\dagger }U_{p_{j}}^{\dagger }U_{\phi }^{\dagger }}{d\phi }U_{\phi
}U_{p_{j}}U_{B}\left \vert 0\right \rangle _{a_{v}}\right. .  \tag{B4}
\end{align}

Noticing $\left[ U_{\phi },U_{p_{j}}\right] =0$, i.e., $U_{\phi }$\ and $%
U_{p_{j}}$ are commutative, and inserting the completeness relation of
number state $\sum \left \vert l\right \rangle _{a_{v},a_{v}}\left \langle
l\right \vert =1$, one can obtian%
\begin{align}
\hat{H}_{1_{j}}& =B_{j}^{2}\overset{\infty }{\underset{l=0}{\sum }}\left.
_{a_{v}}\left \langle 0\right \vert \frac{dU_{B}^{\dagger
}U_{p_{j}}^{\dagger }U_{\phi }^{\dagger }}{d\phi }\left \vert l\right
\rangle _{a_{v},a_{v}}\left \langle l\right \vert \frac{dU_{\phi
}U_{p_{j}}U_{B}}{d\phi }\left \vert 0\right \rangle _{a_{v}}\right.  \notag
\\
& =B_{j}^{2}\overset{\infty }{\underset{l=0}{\sum }}\frac{d}{d\phi }\left.
_{a_{v}}\left \langle 0\right \vert U_{B}^{\dagger }U_{\phi }^{\dagger
}\left \vert l\right \rangle _{a_{v}}\right. U_{p_{j}}^{\dagger }U_{p_{j}}%
\frac{d}{d\phi }\left. _{a_{v}}\left \langle l\right \vert U_{\phi
}U_{B}\left \vert 0\right \rangle _{a_{v}}\right.  \notag \\
& =B_{j}^{2}\overset{\infty }{\underset{l=0}{\sum }}\frac{d}{d\phi }\Pi
_{l}^{\dagger }\left( \eta ,\phi \right) U_{p_{j}}^{\dagger }U_{p_{j}}\frac{d%
}{d\phi }\Pi _{l}\left( \eta ,\phi \right) ,  \tag{B5}
\end{align}%
and
\begin{align}
\hat{H}_{2_{j}}& =iB_{j}^{2}\overset{\infty }{\underset{l=0}{\sum }}\left.
_{a_{v}}\left \langle 0\right \vert \frac{dU_{B}^{\dagger
}U_{p_{j}}^{\dagger }U_{\phi }^{\dagger }}{d\phi }\left \vert l\right
\rangle _{a_{v},a_{v}}\left \langle l\right \vert U_{\phi
}U_{p_{j}}U_{B}\left \vert 0\right \rangle _{a_{v}}\right.  \notag \\
& =iB_{j}^{2}\overset{\infty }{\underset{l=0}{\sum }}\left[ \frac{d}{d\phi }%
\Pi _{l}^{\dagger }\left( \eta ,\phi \right) \right] U_{p_{j}}^{\dagger
}U_{p_{j}}\Pi _{l}\left( \eta ,\phi \right) ,  \tag{B6}
\end{align}%
where $\Pi _{l}^{\dagger }\left( \eta ,\phi \right) =[\Pi _{l}\left( \eta
,\phi \right) ]^{\dag }$ and
\begin{align}
\Pi _{l}\left( \eta ,\phi \right) & =\left. _{a_{v}}\left \langle l\right
\vert U_{\phi }U_{B}\left \vert 0\right \rangle _{a_{v}}\right.  \notag \\
& =\sqrt{\frac{\left( 1-\eta \right) ^{l}}{l!}}e^{i\phi n}\eta ^{\frac{n}{2}%
}a^{l}.  \tag{B7}
\end{align}%
Here, $\Pi _{l}\left( \eta ,\phi \right) $ is actually the Kraus operator,
describing the photon-losses, and satisfying $\sum \Pi _{l}^{\dagger }\left(
\eta ,\phi \right) \Pi _{l}\left( \eta ,\phi \right) =1$, and $n=a^{\dag }a$
is the number operator. $\eta $ is related to the dissipation factor with $%
\eta =1$ and $\eta =0$ being the cases of complete lossless and absorption,
respectively.

For a pure state in extended space, the quantum Fisher information $%
C_{Q_{j}} $ about the parameter $\phi $, is larger or equal to the quantum
Fisher information $F_{L_{j}}$ for mixed state, i.e., $F_{L_{j}}\leq
C_{Q_{j}}$. $C_{Q_{j}}$ is the quantum expression for the Fisher in
formation before optimizing over all possible measurements. Following the
spirit of Ref. \cite{e4}, i.e., in an interferometer with photon losses in
one arm, a possible set of Kraus operators describing the process is%
\begin{equation}
\Pi _{l}\left( \eta ,\phi ,\lambda \right) =\sqrt{\frac{\left( 1-\eta
\right) ^{l}}{l!}}e^{i\phi \left( n-\lambda l\right) }\eta ^{\frac{n}{2}%
}a^{l},  \tag{B8}
\end{equation}%
also satisfying $\sum \Pi _{l}^{\dagger }\left( \eta ,\phi ,\lambda \right)
\Pi _{l}\left( \eta ,\phi ,\lambda \right) =1$. Here $\lambda =0$ and $%
\lambda =-1$ represent the photon losses before the phase shifter and after
the phase shifter, respectively. Thus, one can obtain $F_{L_{j}}$ by
optimizing the parameter $\lambda $ corresponding all possible measurements,
i.e.,
\begin{equation}
F_{L_{j}}=\min_{\Pi _{l}\left( \eta ,\phi ,\lambda \right) }C_{Q_{j}}\leq
C_{Q_{j}}.  \tag{B9}
\end{equation}%
In the paper, we shall use Eqs. (B2), (B8), and (B9) to discuss $F_{L_{j}}$
under photon losses by minimizing $C_{Q_{j}}$ over $\lambda $.

Next, we further derive the normailzation factor $B_{j}$. Using Eq. (B1), it
is ready to have
\begin{equation}
B_{j}^{-2}=\overset{\infty }{\underset{l=0}{\sum }}\frac{\left( 1-\eta
\right) ^{l}}{l!}\left \langle \psi \right \vert a^{\dagger l}\eta
^{n}U_{p_{j}}^{\dagger }U_{p_{j}}a^{l}\left \vert \psi \right \rangle .
\tag{B10}
\end{equation}

To obtain the specific expression of $B_{j}^{-2},$ we appeal to the
technique of integrating within an ordered product of operators (IWOP) \cite%
{e6}, i.e., $\eta ^{n}n^{q}=\colon \partial ^{q}/\partial x^{q}\left \{
e^{\left( \eta e^{x}-1\right) n}\right \} |_{x=0}\colon $, where $\colon
\cdot \colon $ indicates the symbol of the normal ordering form, which
further leads to the formula
\begin{align}
& \overset{\infty }{\underset{l=0}{\sum }}\frac{\left( 1-\eta \right) ^{l}}{%
l!}l^{^{p}}a^{\dagger l}\eta ^{n}n^{q}a^{l}  \notag \\
& =\frac{\partial ^{q+p}}{\partial x^{q}\partial y^{p}}\left[ \eta
e^{x}+(1-\eta )e^{y}\right] ^{n}|_{x=y=0}.  \tag{B11}
\end{align}

Then we can obtain the specific forms for $B_{1}$ and $B_{2}$, i.e.,%
\begin{align}
B_{1}& =[1+\left( 3\eta -\eta ^{2}\right) \left \langle \psi \right \vert
n\left \vert \psi \right \rangle +\eta ^{2}\left \langle \psi \right \vert
n^{2}\left \vert \psi \right \rangle ]^{^{-\frac{1}{2}}},  \tag{B12} \\
B_{2}& =[\left( \eta -\eta ^{2}\right) \left \langle \psi \right \vert
n\left \vert \psi \right \rangle +\eta ^{2}\left \langle \psi \right \vert
n^{2}\left \vert \psi \right \rangle ]^{^{-\frac{1}{2}}},  \tag{B13}
\end{align}%
where
\begin{align}
\left \langle \psi \right \vert n\left \vert \psi \right \rangle & =\alpha
^{2}\cosh ^{2}g+\sinh ^{2}g,  \tag{B14} \\
\left \langle \psi \right \vert n^{2}\left \vert \psi \right \rangle &
=\alpha ^{2}\cosh ^{2}g+\sinh ^{2}g+\alpha ^{4}\cosh ^{4}g  \notag \\
& +2\sinh ^{4}g+4\alpha ^{2}\sinh ^{2}g\cosh ^{2}g,  \tag{B15} \\
\left \langle \psi \right \vert n^{3}\left \vert \psi \right \rangle &
=\alpha ^{2}\cosh ^{2}g+\sinh ^{2}g+3\alpha ^{4}\cosh ^{4}g  \notag \\
& +6\sinh ^{4}g+12\alpha ^{2}\sinh ^{2}g\cosh ^{2}g  \notag \\
& +\alpha ^{6}\cosh ^{6}g+18\alpha ^{2}\cosh ^{2}g\sinh ^{4}g  \notag \\
& +6\sinh ^{6}g+9\alpha ^{4}\cosh ^{4}g\sinh ^{2}g,  \tag{B16}
\end{align}%
and%
\begin{align}
\left \langle \psi \right \vert n^{4}\left \vert \psi \right \rangle &
=\alpha ^{2}\cosh ^{2}g+\sinh ^{2}g+7\alpha ^{4}\cosh ^{4}g  \notag \\
& +14\sinh ^{4}g+28\alpha ^{2}\sinh ^{2}g\cosh ^{2}g  \notag \\
& +36\sinh ^{6}g+6\alpha ^{6}\cosh ^{6}g+24\sinh ^{8}g  \notag \\
& +108\alpha ^{2}\cosh ^{2}g\sinh ^{4}g+\alpha ^{8}\cosh ^{8}g  \notag \\
& +54\alpha ^{4}\cosh ^{4}g\sinh ^{2}g+96\alpha ^{2}\cosh ^{2}g\sinh ^{6}g
\notag \\
& +72\alpha ^{4}\cosh ^{4}g\sinh ^{4}g+16\alpha ^{6}\cosh ^{6}g\sinh ^{2}g.
\tag{B17}
\end{align}%
Here $\langle \cdot \rangle $ is the average under the state $\left \vert
\psi \right \rangle $, and $\left \vert \psi \right \rangle
=U_{S_{1}}\left
\vert \alpha \right \rangle _{a}\left \vert 0\right \rangle
_{b}$ is the state after the first OPA.

Finally,\ using Eq. (B11) and Eqs. (B2), (B8), and (B9) to derive $C_{Q_{j}}$
depending on $\lambda $ for the PA-then-PS ($C_{Q_{1}}$) and for the
PS-then-PA ($C_{Q_{2}}$), we have\
\begin{align}
C_{Q_{1}}& =4\{B_{1}^{2}(u_{1}\left \langle \psi \right \vert n^{4}\left
\vert \psi \right \rangle +u_{2}\left \langle \psi \right \vert n^{3}\left
\vert \psi \right \rangle  \notag \\
& +u_{3}\left \langle \psi \right \vert n^{2}\left \vert \psi \right \rangle
+u_{4}\left \langle \psi \right \vert n\left \vert \psi \right \rangle )
\notag \\
& -[B_{1}^{2}(u_{5}\left \langle \psi \right \vert n^{3}\left \vert \psi
\right \rangle +u_{6}\left \langle \psi \right \vert n^{2}\left \vert \psi
\right \rangle  \notag \\
& +u_{7}\left \langle \psi \right \vert n\left \vert \psi \right \rangle
)]^{2}\},  \tag{B18}
\end{align}%
and

\begin{align}
C_{Q_{2}}& =4\{B_{2}^{2}(u_{1}\left \langle \psi \right \vert n^{4}\left
\vert \psi \right \rangle +u_{8}\left \langle \psi \right \vert n^{3}\left
\vert \psi \right \rangle  \notag \\
& +u_{9}\left \langle \psi \right \vert n^{2}\left \vert \psi \right \rangle
+u_{10}\left \langle \psi \right \vert n\left \vert \psi \right \rangle )
\notag \\
& -[B_{2}^{2}(u_{5}\left \langle \psi \right \vert n^{3}\left \vert \psi
\right \rangle +u_{11}\left \langle \psi \right \vert n^{2}\left \vert \psi
\right \rangle  \notag \\
& +u_{12}\left \langle \psi \right \vert n\left \vert \psi \right \rangle
)]^{2}\},  \tag{B19}
\end{align}%
where%
\begin{align}
u_{1}& =\lambda ^{2}\eta ^{4}-2\lambda ^{2}\eta ^{3}+\lambda ^{2}\eta
^{2}+2\lambda \eta ^{4}-2\lambda \eta ^{3}+\eta ^{4},  \tag{B20} \\
u_{2}& =-6\lambda ^{2}\eta ^{4}+14\lambda ^{2}\eta ^{3}-11\lambda ^{2}\eta
^{2}+3\lambda ^{2}\eta  \notag \\
& -12\lambda \eta ^{4}+22\lambda \eta ^{3}-10\lambda \eta ^{2}-6\eta
^{4}+8\eta ^{3},  \tag{B21} \\
u_{3}& =11\lambda ^{2}\eta ^{4}-28\lambda ^{2}\eta ^{3}+24\lambda ^{2}\eta
^{2}-8\lambda ^{2}\eta  \notag \\
& +\lambda ^{2}+22\lambda \eta ^{4}-52\lambda \eta ^{3}+38\lambda \eta ^{2}
\notag \\
& -8\lambda \eta +11\eta ^{4}-24\eta ^{3}+14\eta ^{2},  \tag{B22} \\
u_{4}& =-6\lambda ^{2}\eta ^{4}+16\lambda ^{2}\eta ^{3}-14\lambda ^{2}\eta
^{2}+4\lambda ^{2}\eta  \notag \\
& -12\lambda \eta ^{4}+32\lambda \eta ^{3}-28\lambda \eta ^{2}+8\lambda \eta
\notag \\
& -6\eta ^{4}+16\eta ^{3}-14\eta ^{2}+4\eta ,  \tag{B23}
\end{align}%
and%
\begin{align}
u_{5}& =\lambda \eta ^{3}-\lambda \eta ^{2}+\eta ^{3},  \tag{B24} \\
u_{6}& =6\lambda \eta ^{2}-3\lambda \eta -3\lambda \eta ^{3}+5\eta
^{2}-3\eta ^{3},  \tag{B25} \\
u_{7}& =4\eta -\lambda +4\lambda \eta -5\lambda \eta ^{2}+2\lambda \eta
^{3}-5\eta ^{2}+2\eta ^{3},  \tag{B26} \\
u_{8}& =-6\lambda ^{2}\eta ^{4}+12\lambda ^{2}\eta ^{3}-7\lambda ^{2}\eta
^{2}+\lambda ^{2}\eta  \notag \\
& -12\lambda \eta ^{4}+18\lambda \eta ^{3}-6\lambda \eta ^{2}-6\eta
^{4}+6\eta ^{3},  \tag{B27}
\end{align}%
as well as%
\begin{align}
u_{9}& =11\lambda ^{2}\eta ^{4}-22\lambda ^{2}\eta ^{3}+13\lambda ^{2}\eta
^{2}-2\lambda ^{2}\eta  \notag \\
& +22\lambda \eta ^{4}-40\lambda \eta ^{3}+20\lambda \eta ^{2}  \notag \\
& -2\lambda \eta +11\eta ^{4}-18\eta ^{3}+7\eta ^{2},  \tag{B28} \\
u_{10}& =-6\lambda ^{2}\eta ^{4}+12\lambda ^{2}\eta ^{3}-7\lambda ^{2}\eta
^{2}+\lambda ^{2}\eta  \notag \\
& -12\lambda \eta ^{4}+24\lambda \eta ^{3}-14\lambda \eta ^{2}  \notag \\
& +2\lambda \eta -6\eta ^{4}+12\eta ^{3}-7\eta ^{2}+\eta ,  \tag{B29} \\
u_{11}& =4\lambda \eta ^{2}-\lambda \eta -3\lambda \eta ^{3}+3\eta
^{2}-3\eta ^{3},  \tag{B30} \\
u_{12}& =\eta +\lambda \eta -3\lambda \eta ^{2}+2\lambda \eta ^{3}-3\eta
^{2}+2\eta ^{3}.  \tag{B31}
\end{align}%
Then, we can further optimize $\lambda $ to get the minimum value of $%
C_{Q_{j}}$ using Eq. (\ref{eq21}), which corresponding to the $F_{L_{j}}$%
.\bigskip

\textbf{APPENDIX C: THE WF WITH NCO}\bigskip

For a two-mode quantum state $\rho $, its WF under the coherence state
representation can be calculated as
\begin{align}
W_{j}\left( z,\gamma \right) & =e^{2(\left \vert z\right \vert ^{2}+\left
\vert \gamma \right \vert ^{2})}\int \frac{d^{2}\beta _{a}d^{2}\beta _{b}}{%
\pi ^{4}}  \notag \\
& \times \left \langle -\beta _{a},\beta _{b}\right \vert \rho \left \vert
\beta _{a},\beta _{b}\right \rangle  \notag \\
& \times e^{2\left( z\beta _{a}^{\ast }-z^{\ast }\beta _{a}+\gamma \beta
_{b}^{\ast }-\gamma ^{\ast }\beta _{b}\right) },  \tag{C1}
\end{align}%
where $j=1$ or $2$, and $\left \vert \beta _{a},\beta _{b}\right \rangle
=\left \vert \beta _{a}\right \rangle \otimes \left \vert \beta
_{b}\right
\rangle $\ are two-mode coherent states. From Eq.(C1), the
analytic expression of the WF can be obtained by providing the density
operator $\rho $\ of the quantum state. Here we only consider the ideal
case, i.e., without losses. The quantum state after the NCO is $\left \vert
\psi _{P}\right
\rangle =A_{j}U_{P_{j}}U_{S_{1}}\left \vert \alpha
\right
\rangle _{a}\left
\vert 0\right \rangle _{b}.$\ Therefore, the
density operator $\rho $\ can be expressed as
\begin{equation}
\rho =\left \vert \psi _{P}\right \rangle \left \langle \psi _{P}\right
\vert .  \tag{C2}
\end{equation}%
By substituting Eq. (C2) into Eq. (C1), we can obtain the WF after the NCO.
Here the specific expressions are not shown for simplicity.

To clearly observe the effect of the gain factor $g$\ on the nonclassicality
of two different non-Gaussian operations ($aa^{\dagger }$, $a^{\dagger }a$),
we use the negative volume $V_{j}$\ of the WF to quantitatively describe the
non-classicality of the quantum state after the NCO. The calculation formula
for the negative volume $V_{j}$\ of the WF is given by
\begin{equation}
V_{j}=\frac{\int dx_{1}dx_{2}dy_{1}dy_{2}\left[ \left \vert W_{j}\left(
z,\gamma \right) \right \vert -W_{j}\left( z,\gamma \right) \right] }{2},
\tag{C3}
\end{equation}%
where $z=\left( x_{1}+iy_{1}\right) /\sqrt{2},\gamma =\left(
x_{2}+iy_{2}\right) /\sqrt{2}.$\ According to Eq. (C3), the WF negative
volume of the state $\left \vert \psi _{P}\right \rangle $\ can be
numerically calculated.

\end{document}